\documentclass[amsfonts,amssymb,amsmath,eqsecnum]{revtex4-1}

\usepackage{cancel}
\usepackage{bm}
\usepackage{graphicx}
\usepackage{color}

\newcommand{\stl}[1]{ \mbox{ $ \hspace{0.1em} \stackrel{ \rule{0.3pt}{0.275ex} \hspace{-0.25pt} \overline{\hphantom{\mbox{$\displaystyle #1$}}} \hspace{-0.30pt} \rule{0.3pt}{0.275ex} }{#1} \hspace{0.2em} $ } }

\begin{document}
\title{Feedback control of flow alignment in sheared liquid crystals}
\author{David~A.~Strehober}
\email{physik@strehober.de}
\author{Eckehard Sch\"oll}
\author{Sabine~H.~L. Klapp}
\email{klapp@physik.tu-berlin.de}
\affiliation{Institut f\"ur Theoretische Physik, Sekr. EW 7-1, Technische Universit\"at Berlin,
Hardenbergstra{\ss}e 36, D-10623 Berlin, Germany}

\begin{abstract}
Based on a continuum theory, we investigate the manipulation of the non-equilibrium behavior of a sheared liquid crystal via closed-loop feedback
control. Our goal is to stabilize a specific dynamical state, that is, the stationary "flow-alignment",  under conditions where the uncontrolled system displays oscillatory director dynamics with in-plane symmetry.
To this end we employ time-delayed feedback
control (TDFC), where the equation of motion for the $i$th component, $q_i(t)$, of the order parameter tensor is supplemented by a control term involving the difference 
$q_i(t)-q_i(t-\tau)$. In this diagonal scheme, $\tau$ is the delay time.
We demonstrate that the TDFC method successfully stabilizes flow alignment for suitable values of the control strength, $K$, and $\tau$; these values are determined by solving an exact eigenvalue equation. Moreover, our results show that only small values of $K$ are needed when the system is sheared from an isotropic equilibrium state, contrary to the case where the equilibrium state is nematic.
\end{abstract}

\maketitle

\section{Introduction}
\label{intro}
Liquid crystals under shear can display a variety of non-equilibrium dynamical states determining the motion of the director of the
(shear-induced or spontaneous) orientational ordering. The simplest of these states is the stationary "flow-alignment" typically occurring at large shear rates and/or large values of the
(particle geometry-related) coupling parameter $\lambda_{\mathrm{K}}$. However, the systems can also display various types of oscillatory motion, spatio-temporal symmetry breaking,
and even chaotic behavior \cite{Rien02,Rien02a,Grosso03,Forest04,Ripoll08,Das05}.  
The discovery of this rich dynamical behavior has stimulated intense research both by theoretical methods
(such as continuum approaches \cite{Hess75,Hess76b,Doi80,Doi81,Olmsted92}
and particle-based computer simulations \cite{Tao05,Tao09,GERM05,Ripoll08}) and by experiments (see, e.g., \cite{Sood00,Lettinga05}). 

The nonlinear orientational dynamics also has direct implications for
the rheological behavior of the system as reflected, e.g., by non-monotonic
stress-strain curves ("constitutive relations")
\cite{FIEL07,ARA05,Cates02,Aradian06,GOD08,Klapp10,Heidenreich09}
and a non-Newtonian behavior of the viscosity.
Understanding the dynamics is thus a prerequisite for the deliberate design of materials with specific rheological properties,
which are tunable by parameters such as particle geometry, concentration (temperature) and external fields.

Beyond pure understanding, however, one may wish to {\it stabilize} a certain dynamic state with a well-defined associated rheology. A candidate for stabilization could
be the stationary shear-alignment state. Indeed, it has been shown
\cite{Hess04,Klapp10} that the viscosity in such a state is particularly low ("shear-thinning"),
in fact, lower than the viscosity of the corresponding unsheared
system. In other words the stationary alignment of the liquid crystal molecule in the shear flow tends to lower frictional effects. This situation changes dramatically when the nematic director
starts to oscillate \cite{Klapp10}. Thus, shear-aligned
systems may serve as particularly good lubricants.

In this paper we investigate the possibility to stabilize the flow-aligned state 
by a continuum approach
for the orientational dynamics \cite{Hess75,Hess76b}, 
combined with the method of time-delayed feedback control (TDFC). The relevant dynamical variable
within the continuum approach is the second-rank alignment tensor ${\bf a}(t)$ carrying five independent components.
In a previous, short study \cite{Strehober2012} we have already shown the TDFC to be successful if the dynamics of the full
${\bf a}(t)$ is simplified into that of a two-dimensional director characterizing uniaxial, shear-induced ordering within the shear plane.
In the present paper we release this somewhat artificial restriction and investigate the full (in-plane) dynamics under TDFC. 
We focus on conditions where the uncontrolled system displays a wagging-like oscillatory motion within the shear plane.
In this situation it seems tempting to consider only a reduced (three-dimensional system) involving only
those components of the order parameter ${\bf a}$, which describe in-plane dynamics. However, as it was demonstrated in previous studies \cite{Rienacker2000,Rien02a}, this reduction can predict a stable fixed point (the so-called log-rolling) 
which is actually unstable after inclusion of the remaining components of the order parameter.
We thus consider the full five dimensional dynamical system to explore stability.
\\
As in our earlier work \cite{Strehober2013a}, we did make some simplifying assumptions.
First, we do not consider back-coupling of the orientational dynamics onto the flow. Rather we assume that the velocity field is imposed externally.
Extensions of the theory incorporating such effects are proposed in a study of Lima and Rey \cite{Lima2004} as well as by Heidenreich \emph{et al.} \cite{Heidenreich09}.
Second we neglect the role of boundaries, which was discussed by Tsuji and Rey \cite{Tsuji1997} as well as in \cite{Heidenreich09}.
A third assumption is that we consider our sheared system being free of defects (for corresponding extensions see \cite{Rey2002}).
\\
Moreover, in the context of TDFC, we explore the interplay between the performance of the control scheme, on the one side, and the nature of the underlying
equilibrium phase from which the liquid crystal is sheared, on the other side.
TDFC is a closed-loop control method proposed 1992 by Pyragas \cite{Pyragas92}, which allows one to
stabilize periodic and steady states which would be unstable otherwise.    
In the meantime, TDFC has been applied to a broad variety of nonlinear
systems including semiconductor nanostructures \cite{BAB02,UNK03,SCH03a,KEH09},
lasers \cite{DAH08b,DAH10}, excitable media \cite{SCH06c,Schlesner08,KYR09}, and neural
systems \cite{DAH08,SCH08,PAN12}
(see \cite{Schoellbuch,Schoellreview} for
overviews).
Within the Pyragas method, the equations of motion are supplemented by control terms built on the differences
$q_i(t)-q_i(t-\tau)$ between the present and an earlier value of an appropriate system variable $q_i$. 
This type of control is noninvasive as the control forces vanish when the steady state (or a periodic state
with period $T=n \tau$, with $n=1,2,3,\dots$) is reached. 
A general, analytic investigation of the
application of TDFC to steady states has been given in
Ref.~\cite{HOV05}. 
 
The paper is organized as follows. We start 
in Sec.~\ref{Dynamic} with a
review of the basic dynamical equations for the order parameter, $\bm{a}(t)$. 
In Sec.~\ref{in-plane} we summarize the dynamical behavior occurring
in dependence of the shear rate and the coupling parameter for two temperatures (and correspondingly, different phases) of the underlying equilibrium system
(for a full discussion based on a bifurcation analysis, see Ref.~\cite{Strehober2013a}. For each reduced temperature $\theta$,
we select a parameter set in which the uncontrolled system displays oscillatory director dynamics within the shear plane.
 The theoretical background of the TDFC method is introduced in Sec.~\ref{delay}. Numerical results for the selected parameter sets
are presented in Sec.~\ref{results}. 
Finally, we give a conclusion and outlook.

\section{Background: Continuum theory of the orientational dynamics under shear}
\label{Dynamic}
We employ a mesoscopic description 
of the system, where the relevant dynamic variable is the orientational order parameter averaged over some volume in space. In a sheared liquid crystal, this order parameter corresponds to the time-dependent, 2nd-rank alignment tensor ${\bf a}=\sqrt{15/2}\langle \stl{{\bf u}{\bf u}}\rangle$, where 
${\bf u}$ describes the orientation of the molecular axis and $\stl{\ldots}$ indicates the symmetric traceless part of a tensor.
The average $\langle\ldots\rangle$ is defined as (see Ref.~\cite{Strehober2013a})
\begin{equation}
\label{eq:average}
\langle\ldots\rangle = \int_{S^2}{\mathrm{d}}^2\bm{u}\,\ldots \,\rho^{\mathrm{or}}(\bm{u},\bm{r},t),
\end{equation}
involving the orientational distribution function $\rho^{\mathrm{o}r}(\bm{u},\bm{r},t)$ \cite{Sonnet1995}. The integral in Eq.~(\ref{eq:average})
is performed over the unit sphere. The orientational distribution is defined as
$\rho^{\mathrm{or}}(\bm{u},\bm{r},t) = N^{-1} \langle \sum_{i=1}^N \delta(\bm{u}-\bm{u}_i(t))\rangle_{\mathrm{ens}}$, where $\bm{u}_i$ is the microscopic orientation of particle $i$ ($i=1,\ldots,N$),
and $\langle\ldots\rangle_{\mathrm{ens}}$ is an ensemble average 
in a small volume $dV$ around the space point $\bm{r}$ at time $t$. 

In the isotropic equilibrium state, all components of ${\bf a}$ are zero, whereas nematic 
ordering (which may be uniaxial or biaxial in character) is characterized by one or several components of  ${\bf a}$ being non-zero.

Switching on an external shear flow characterized by a velocity field ${\bf v}$, the alignment tensor becomes a time-dependent quantity.
Its equation of motion can be derived from a generalized Fokker-Planck equation \cite{Hess76b,Doi80,Doi81}
or, alternatively, from irreversible thermodynamics \cite{Hess75},
yielding for a homogeneous system (in dimensionless form) \cite{Grandner07}
\begin{equation}
\label{eq:Hess}
\frac{d {\bf a}}{d t}=2\stl{{\bm \Omega}\cdot{\bf a}}
+2\sigma\,\stl{{\bm \Gamma}\cdot{\bf a}}-{\Phi}'(\bm{a})
+\sqrt{\frac{3}{2}}\lambda_{\mathrm{K}}\,{\bm \Gamma}.
\end{equation}

In Eq.~(\ref{eq:Hess}), 
${\bm \Gamma}=\left((\nabla {\bf v})^T+\nabla{\bf v}\right)/2$ is the strain rate tensor (with the superscript "T" denoting the transpose
of tensor $\nabla{\bf v}$) and
${\bm \Omega}=\left((\nabla {\bf v})^T-\nabla{\bf v}\right)/2$ is the vorticity. 
The symbol $\stl{\bf{x}}$ indicates the symmetric traceless part of a tensor $\bf{x}$, i.e. $\stl{{\bf x}} = 1/2 ({\bf x}+{\bf x}^T) - 1/3 Tr({\bf x})$.
In the present work we consider a planar Couette flow
characterized by ${\bf v}=\dot\gamma y {\bf e}_{\mathrm{x}}$, with $\dot\gamma$ being the shear rate and ${\bf e}_{\mathrm{x}}$ being a unit vector.
This yields ${\bm \Gamma}=\dot\gamma\stl{{\bf e}_{\mathrm{x}}{\bf e}_{\mathrm{y}}}$ and 
${\bm \Omega}=(\dot\gamma/2) \left({\bf e}_{\mathrm{x}}{\bf e}_{\mathrm{y}}-{\bf e}_{\mathrm{y}}{\bf e}_{\mathrm{x}}\right)$, respectively. The (dimensionless) parameter 
$\lambda_{\mathrm{K}}$ is the so-called "tumbling" parameter, which
measures the coupling strength between alignment and strain. This
parameter is related to the shape (i.e., the aspect ratio) of the
particles \cite{Hess76b}. The relaxation parameter
$\sigma$ plays only a minor role, and following previous works \cite{Hess2004,Rien02,GrandnerEPJ2007,GrandnerPRE2007,Strehober2013a},
we set $\sigma=0$. 
Finally, the (tensorial) quantity ${\Phi}'(\bm{a})$ appearing in Eq.~(\ref{eq:Hess}) corresponds to the derivative
of the free energy with respect to the (non-conserved) order parameter, i.e., ${\Phi}'(\bm{a})=\partial \Phi/\partial {\bf a}$. 
We employ the (dimensionless) Landau-de Gennes (LG) expression for
the free energy \cite{deGennes} given by
\begin{equation}
\label{free_energy}
 \Phi=\frac{\theta}{2}{\bf a}:{\bf a}-\sqrt{6}\left({\bf a}\cdot{\bf a}\right):{\bf a}+\frac{1}{2}\left({\bf a}:{\bf a}\right)^2,
 \end{equation}
where the notation ":" stands for the trace over the product of two tensors, and "$\cdot$" indicates conventional matrix multiplication. In Eq.~(\ref{free_energy}), $\theta$ 
plays the role of an effective, dimensionless temperature, which is the tuning parameter for the isotropic-nematic transition
in thermotropic liquid-crystal-systems.
A first order isotropic-nematic transition occurs at $\theta=1$.
For temperatures $\theta > 1$ ($\theta < 1$) the isotropic (nematic) phase is stable, i.e., it corresponds to the lowest minimum of the free energy.
Upon ``cooling down'' from high temperatures, the nematic state appears as a metastable phase already at $\theta = 9/8$. Crossing the phase transition (at $\theta=1$), the isotropic phase
remains as a metastable phase down to $\theta=0$, below which it becomes unstable. We note that this general scenario applies not only to thermotropic liquid crystals
(where $\theta$ is related to a true temperature), but also to lyotropic liquid crystals and suspensions of colloidal rods. In these cases, the isotropic-nematic
transition is triggered by the concentration, and $\theta$ has to be defined accordingly \cite{Strehober2013a}, otherwise the approach remains the same.

Equation~(\ref{eq:Hess}) is most conveniently solved by expanding $\bm{a}$
and the other tensors appearing on the right side into a tensorial basis set (see, e.g., \cite{Rien02a}), e.g.,
$\bm{a}=\sum_{l=0}^4 a_l{\bf T}^l$, where $a_l$ are the (five) independent components of ${\bf a}$, and
the (orthonormal) tensors ${\bf T}^l$ 
involve linear combinations of the unit vectors ${\bf e}_{\mathrm{x}}$, ${\bf e}_{\mathrm{y}}$, and ${\bf e}_{\mathrm{z}}$
(for explicit expressions, see e.g. in Ref.~\cite{Rien02a}). One then obtains the five-dimensional dynamical system
\begin{align}
  \dot{\bm{q}} &= \bm{F},
  \label{dyn_system}
\end{align}
where the vector $\bm{q} = (a_0,a_1,a_2,a_3,a_4)$, and the components of the vector $\bm{F}$ are given by
\begin{align}
{F}_0= &-\phi_0\notag\\
{F}_1= &-\phi_1 +\dot{{\gamma}}a_2\notag\\
{F}_2= & -\phi_2 -\dot{{\gamma}}a_1+{\frac{1}{2}}{\sqrt{3}}{\lambda}_K\dot{{\gamma}}\notag\\
{F}_3=  &-\phi_3+{\frac{1}{2}}\dot{{\gamma}}a_4\notag\\
{F}_4= &-\phi_4-{\frac{1}{2}}\dot{{\gamma}}a_3\label{eq:HomSyst1}.
\end{align}
In Eqs.~(\ref{eq:HomSyst1}), the quantities $\phi_l$ ($l=0,\ldots,4$) represent the components of the vector ${\bm \Phi}$ (that consist of the projections of ${\Phi}'(\bm{a})$ on the tensor basis). 
These quantities are nonlinear functions of the $a_l$; explicit expressions are given in the appendix.

\section{Feedback control}
\label{control}
\subsection{Choosing candidate states}
\label{in-plane}
\begin{figure}
\includegraphics[width=\columnwidth]{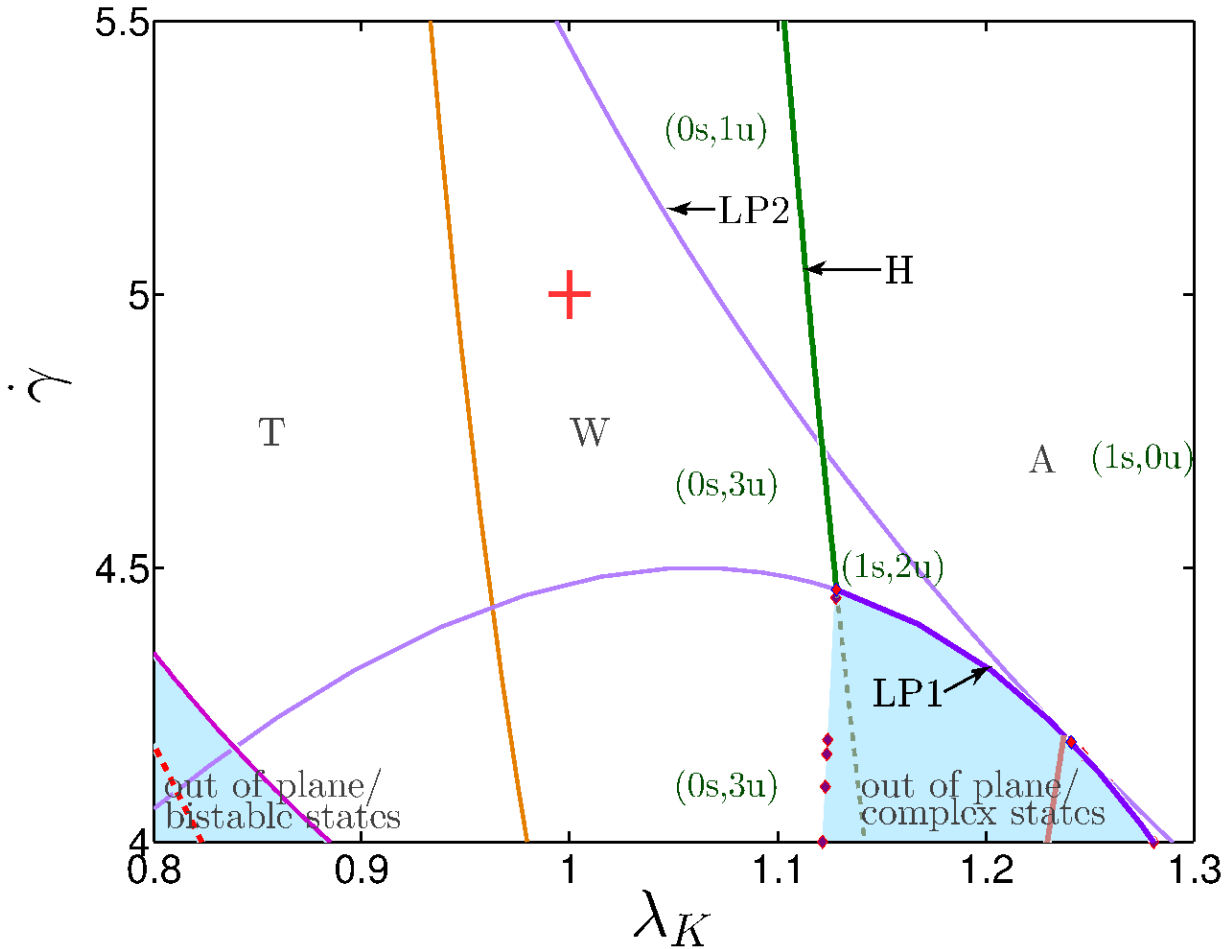}
\caption{\label{fig:parameterset1}
(Color online) Dynamical state diagram in the plane spanned by the tumbling parameter $\lambda_{\mathrm{K}}$ and the
(dimensionless) shear rate $\dot\gamma$ for a nematic system ($\theta=0$).
The lines were obtained via a codimension-2 bifurcation analysis as described in Ref.~\cite{Strehober2013a}.
The nearly vertical line represents a supercritical Hopf bifurcation line (H). On the right side of the H line we find flow alignment (A), corresponding to a
stable fixed point. This is indicated by the notation (1s,0u), where "1s" and "0u" means one stable (s) and no unstable (u) fixed point, respectively. 
Upon crossing the H line towards lower values of $\lambda_{\mathrm{K}}$, oscillatory in-plane states [wagging (W) and tumbling (T)] become stable. 
As indicated by the notations (0s,3u) [and (0s,1u)] there is no stable fixed point in this area, but three [one] unstable ones. 
The cross marks the parameter set $\bm{\beta}_{\mathrm{I}}=(\lambda_{\mathrm{K}}=1.0,\dot{\gamma}=5.0,\theta=0)$ where we apply feedback control.
The shaded areas appearing at low $\dot\gamma$ correspond to oscillatory states with out-of-plane symmetry.
The lines labeled LP1 and LP2 are limit point lines (saddle-node bifurcation lines).}
\end{figure}

The dynamical behavior emerging from the mesoscopic equations of motion (\ref{dyn_system}) has been studied in detail for a variety of temperatures $\theta$
(determining the behavior of the unsheared system)
and a broad range of shear rates $\dot\gamma$ and shear coupling parameters $\lambda_{\mathrm{K}}$ \cite{Strehober2013a,Hess2004,Heidenreich2008Thesis,Rienacker2000,Rien02,Rien02a}.
In particular in Ref.~\cite{Strehober2013a}, we have investigated systems at different temperatures via a numerical bifurcation analysis (for numerical details see Appendix of \cite{Strehober2013a}).
Special attention has been devoted to systems sheared from the stable or metastable nematic equilibrium phase ($\theta\leq 9/8 $). 
An exemplary dynamical state diagram for the case $\theta=0$ is shown in Fig.~\ref{fig:parameterset1} (data taken from Ref.~\cite{Strehober2013a} and the diagram is consistent with earlier works, i.e. \cite{Rien02,Rien02a,Rienacker2000}).

\begin{figure}
\includegraphics[width=\columnwidth]{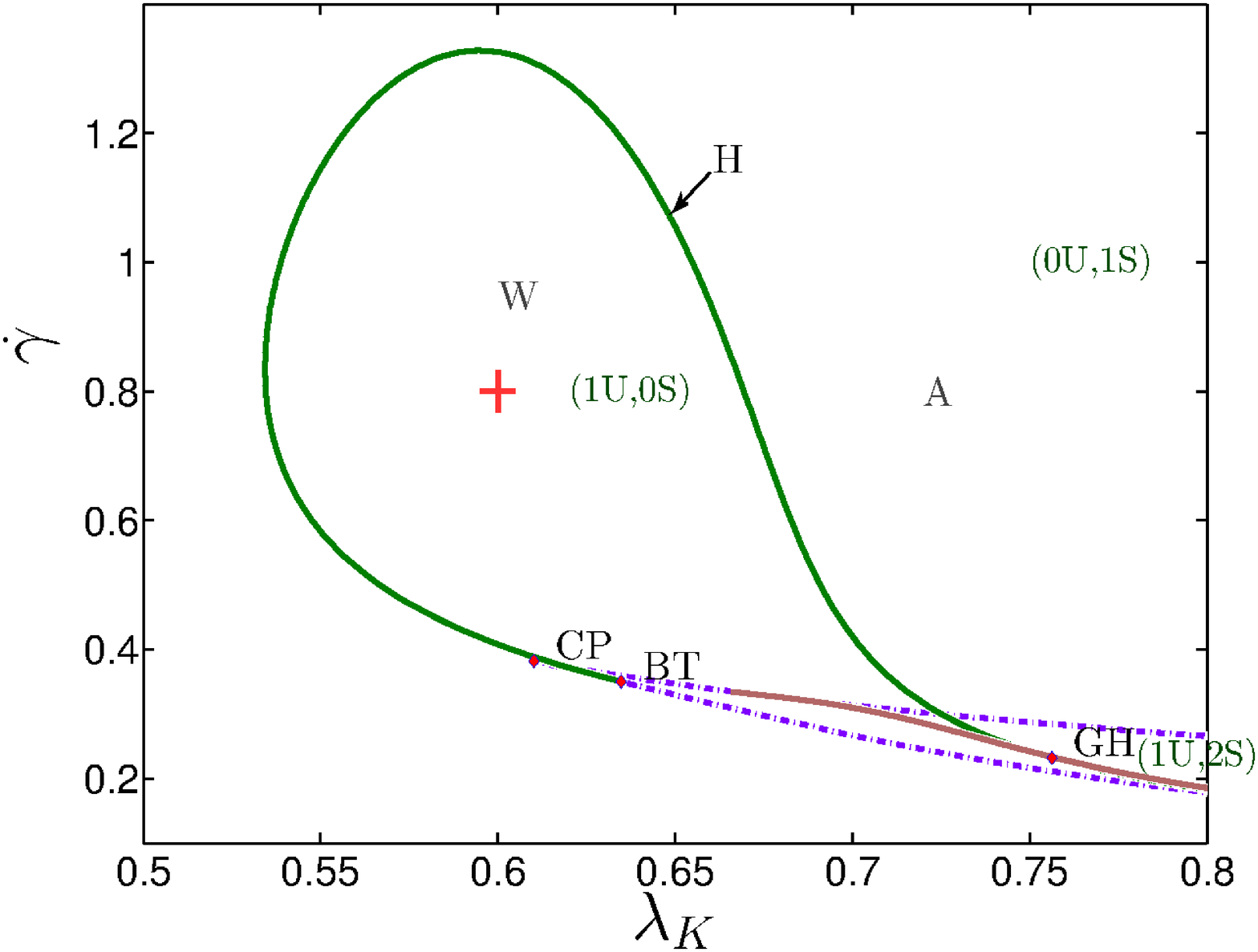}
\caption{\label{fig:parameterset2}
(Color online) Dynamical state diagram at $\theta=1.20$, where the unsheared system is isotropic. The lines are obtained via a codimension-2 bifurcation analysis
as described in Ref.~\cite{Strehober2013a}. With the bifurcation analysis we also detect special points like cusp point (CP), Bogdanov-Takens point (BT) and generalized Hopf point (GH).
The solid line (shaped like an oval) represents a supercritical H bifurcation line, within which the system is in an oscillatory (W) state. 
The cross marks the parameter set $\bm{\beta}_{\mathrm{II}}=(\lambda_K=0.6,\dot{\gamma}=0.8,\theta=1.20)$ selected as a second candidate for feedback control.
Outside this W regime
(characterized by one unstable fixed point), the system is in a stationary, flow-aligned state (A). The dashed lines appearing in the right, lower corner of the diagram
are limit point lines (LP) (saddle-node bifurcation lines). Between the LP curves one finds areas of bistability between paranematic and nematic states (for details on that see Ref.~\cite{Strehober2013a}). The notation
regarding the fixed points is as in Fig.~\ref{fig:parameterset1}.}
\end{figure}
For large values of the shear rate ($\dot\gamma\gtrsim 4.5$, i.e., above the semicircle line in Fig.~\ref{fig:parameterset1}) the stable dynamical states have in-plane symmetry. In this case the main director is restricted to directions within the shear plane (i.e., the $x$-$y$-plane), implying 
that the components $a_3=a_4=0$. 
At large values of the coupling parameter $\lambda_{\mathrm{K}}$, this in-plane state is stationary in character, reflecting that the nematic director is "frozen" and encloses a fixed angle
with the direction of the shear flow. This is the so-called "flow alignment" state, which we label by A.  
Mathematically, the A state corresponds to a stable fixed point of the dynamics. Decreasing the coupling parameter $\lambda_{\mathrm{K}}$
(at fixed, large $\dot\gamma$), one encounters 
a supercritical Hopf bifurcation \cite{Strehober2013a} and the system displays oscillatory states with in-plane symmetry. These are
the so-called wagging (W) occurring at intermediate values of $\lambda_{\mathrm{K}}$, and the tumbling (T) state at low $\lambda_{\mathrm{K}}$,
Both T and W are characterized by the presence of stable limit cycles, and, correspondingly, unstable fixed points (for a more detailed discussion, see 
Sec.~\ref{results}). In the W state the angle between the nematic director
and the flow direction oscillates periodically between a minimal and a maximal
value, whereas in the T state, the director performs full, in-plane rotations. We
stress, however, that there is no fundamental difference
between W and T motion in the sense that these states are not separated by a bifurcation \cite{Strehober2013a}.

At lower shear rates, additional dynamical states appear which are characterized by non-zero values
of all five components of the order parameter. Physically, this means that the main director 
performs oscillations not only within the shear plane, but also
out of this plane. Typical representatives 
are the "kayaking wagging" (KW) and "kayaking tumbling" (KT) states first observed in Ref.~\cite{Larson1991}. 
In Fig.~\ref{fig:parameterset1}, such out-of-plane solutions occur in the shaded regions. 
We have also indicated regions of bistability and complex chaotic behavior (for a more detailed discussion, see Refs.~\cite{Strehober2013a,Rienacker2000,Rien02,Rien02a}).

The main goal of the present paper is to explore the stabilization
of the fixed point corresponding to flow alignment  within a parameter range where the system is in an {\it in-plane} oscillatory state (i.e., W or T). 
The specific position in parameter space is indicated by the cross in Fig.~\ref{fig:parameterset1}, corresponding to the parameter set $\bm{\beta}_{\mathrm{I}}:=(\lambda_K=1.0,\dot{\gamma}=5.0,\theta=0)$.

Our reasoning to focus on in-plane situations is twofold: First, the absence of stable out-of-plane solutions allows us to focus on
only three components of the order parameter tensor, that is, $a_0$, $a_1$ and $a_2$. Second, it has been shown \cite{Strehober2013a} that the out-of-plane states 
do {\em not} arise via a Hopf bifurcation. Thus, there is no "natural" unstable fixed point which one could try to stabilize via TDFC. 

In addition to a nematic system, we also consider in our study a system which is {\it isotropic} in the absence of shear ($\theta=1.20$). 
A corresponding state diagram is shown in Fig.~\ref{fig:parameterset2}.
%
%
At large coupling parameters ($\lambda_{\mathrm{K}}\gtrsim 0.75$), the shear flow induces first ($\dot\gamma\lesssim 0.2$)
a "paranematic" ordering characterized by very small values of the order parameters. Increasing $\dot\gamma$ the systems then transforms via a first-order transition
into a flow-aligned state (A). Interestingly, however, the system can also display an in-plane oscillatory state, that is, wagging (W). This behavior is rather surprising in view of the isotropic nature of the underlying equilibrium state and was detected only recently \cite{Strehober2013a}. As seen from Fig.~\ref{fig:parameterset2}, the wagging occurs in a parameter island located at lower values of $\lambda_{\mathrm{K}}$. The major part of the boundaries of this island represent Hopf bifurcations. We select a parameter point within this
wagging island as a second candidate for feedback control. Specifically, $\bm{\beta}_{\mathrm{II}}:=(\lambda_K=0.6,\dot{\gamma}=0.8,\theta=1.20)$.

Having identified suitable parameter sets ($\bm{\beta}_{\mathrm{I}},\bm{\beta}_{\mathrm{II}}$) to apply feedback control of steady states, we now turn to a detailed discussion of (i) the stability of the corresponding steady states, and (ii) their behavior under TDFC with diagonal control scheme. The corresponding methods are outlined in the subsequent
Sec.~\ref{delay}. In Sec.~\ref{results}, we will present the numerical results.

As already remarked in the introduction, we perform the stability analysis described below with the full, five-dimensional system (see also discussion in Sec.~\ref{stabilization_nematic}).
In this way we avoid difficulties arising if one considers the three-component system alone.
\subsection{Time-delayed feedback control}
\label{delay}
As a starting point for feedback control, we first need to determine the steady states (fixed points) $\bm{q}^\star=(a_0^\star,a_1^\star,a_2^\star,a_3^\star,a_4^\star)$ 
of the dynamical system [see Eq.~(\ref{dyn_system})] corresponding to the two parameter sets $\bm{\beta}_{\mathrm{I}}$, $\bm{\beta}_{\mathrm{II}}$. 
The fixed points fulfill the condition
$\dot{\bm{q}^{\star}}=0$. We have solved these equations numerically.

For each fixed point, its linear stability can be checked by considering the $5\times 5$ Jacobian ${\bf J}$ of the dynamical system. 
The elements 
$J_{ij}=\partial F_i/\partial q_j$ of ${\bf J}$
are given in the appendix \ref{appA}.
Small perturbations $\delta{\bf q}(t)$ away from the steady state evolve with time as $\dot{\delta{\bf q}}(t)={\bf J}\delta{\bf q}(t)$. This linear equation can be solved with the ansatz
$\delta{\bf q}(t)={\bf A}\exp[\nu t]$ (with ${\bf A}$ containing the real amplitudes of the perturbation), yielding the eigenvalue 
equation 
\begin{equation}
\label{eigen_nocontrol}
\nu \delta{\bf q}={\bf J}\delta{\bf q}.
\end{equation}
The eigenvalues can then be calculated from the characteristic equation
$\mathrm{det}\left({\bf J}-\nu{\bf I}\right)=0$ (where ${\bf I}$ is the unity matrix).
Stability of the fixed point $\bm{q}^\star=(a_0^\star,a_1^\star,a_2^\star,a_3^\star,a_4^\star)$ requires that all eigenvalues of $\bm{J}$ evaluated at this fixed point have negative real parts, implying that perturbations die off with time.

We now aim to stabilize the unstable fixed point ${\bf q}^\star$ corresponding to flow alignment 
within the range, where the system ends up in wagging motion. To this end, we use the TDFC method \cite{Pyragas92}. 
Following earlier work \cite{HOV05}, we employ a diagonal 
control scheme, where the control force acting on the $i$th component 
(with $i=0,1,2,3,4$) involves only the same component. Explicitly,
%
%
\begin{align}
\label{eq:control}
\dot{q_i}=F_i({\bf q},\bm{\beta})-K\left(q_i(t)-q_i(t-\tau)\right)
\end{align}
where $K$ measures the strength of control and $\tau$ is the delay time. Note that the feedback terms in Eq.~(\ref{eq:control}) vanish when the fixed point is fully stabilized, that is,
if ${\bf q}^\star(t-\tau)={\bf q}^{*}(t)$.
The impact of the control on the phase portrait, for the two different parameter sets ($\bm{\beta}_{\mathrm{A}}$, $\bm{\beta}_{\mathrm{II}}$), is shown in Sec.~\ref{results}.

To get a better insight into the role of the feedback control, it is instructive to perform a (linear) stability analysis of the delayed differential equations given in (\ref{eq:control})
\cite{HOV05}. In analogy to the procedure discussed before [see Eq.~(\ref{J_original} below)], we consider
a small displacement from the fixed point, $\delta{\bf q}(t)$. To linear order, the dynamics of this displacement follows from Eq.~(\ref{eq:control}) as 
\begin{equation}
\label{delta_x_der}
\dot{\delta{\bf q}}=\left({\bf J}-K{\bf I}\right)\delta{\bf q}(t)+K\delta{\bf q}(t-\tau).
\end{equation}
%

This equation can be solved with the exponential ansatz $\delta{\bf q}(t)={\bf B}\exp[\mu t]$, where ${\bf B}$ contains the amplitudes
of the displacement, and $\mu$ is a complex number. Inserting this ansatz into Eq.~(\ref{delta_x_der}) one obtains the eigenvalue equation
\begin{equation}
\label{eigen_control}
\left(\mu+K\left(1-\exp[-\mu\tau]\right)\right)\delta{\bf q}={\bf J}\delta{\bf q}.
\end{equation}
The corresponding characteristic equation yielding the eigenvalues $\mu$ is given by
\begin{equation}
\label{control_characteristic}
\mathrm{det}\left({\bf J}-\left(\mu+K\left(1-\exp[-\mu\tau]\right)\right){\bf I}\right)=0.
\end{equation}

However, an even simpler way to calculate the eigenvalues $\mu$ is based on the following notion:
Equation~(\ref{eigen_control}) has {\it exactly} the same form as the corresponding equation for the uncontrolled case, Eq.~(\ref{eigen_nocontrol}). 
In other words, the linear operator ${\bf J}$  has the same set of eigenfunctions $\delta{\bf q}$ in both, the uncontrolled and the controlled case.
This notion implies that if the eigenvalues $\nu$
of the uncontrolled system are known, then the eigenvalues $\mu$ of the controlled system can be calculated from 
\begin{equation}
\label{eigen_equation}
\mu+K\left(1-\exp[-\mu\tau]\right)=\nu
\end{equation}
 We also stress that Eq.~(\ref{eigen_equation}) is {\it equivalent} to the eigenvalue equation derived in Ref.~\cite{HOV05}. In that paper, the (diagonal) feedback control of unstable steady states
 in a two-dimensional dynamical system
 was studied from a general perspective, that is, without reference to a particular physical system. In particular, it was shown that Eq.~(\ref{eigen_equation}) can be solved {\it analytically} by using the Lambert function ${\cal W}$ \cite{Corless96}. 
 The same strategy can be used in the present, five-dimensional case, because we are using the same, diagonal control scheme.
 To see this, we rearrange Eq.~(\ref{eigen_equation}) into
 \begin{equation}
\label{eigen_equation_2}
\left(\mu+K-\nu\right)\tau=K\tau\exp[-\mu\tau].
\end{equation}
 Setting $z=\left(\mu+K-\nu\right)\tau$ and multiplying both sides of Eq.~(\ref{eigen_equation_2}) with $\exp[z]$ we have
 \begin{equation}
\label{eigen_equation_3}
z\exp[z]=K\tau\exp[\left(K-\nu\right)\tau]\equiv g.
\end{equation}
We can solve this equation with respect to $z$ by using that $z={\cal W}(z\exp[z])={\cal W}(g)$. After re-substituting (i.e., $\mu\tau=z-(K-\nu)\tau$) we finally obtain the explicit formula
 \begin{equation}
 \label{lambert}
 \mu\tau={\cal W}\left(K\tau\exp[-\nu\tau+K\tau]\right)+\nu\tau-K\tau.
 \end{equation}
We have calculated the eigenvalues both, numerically [from Eq.~(\ref{control_characteristic})]
and analytically [from Eq.~(\ref{lambert})] for a range of control parameters $K$, $\tau$, at the two parameter sets
$\bm{\beta}_{\mathrm{I}}$ and $\bm{\beta}_{\mathrm{II}}$ (see Figs.~\ref{fig:parameterset1} and \ref{fig:parameterset2}, respectively).
Notice that the TDFC scheme, which is based on the coupling of a dynamical variable
to its own history [see Eq.~(\ref{eq:control})], creates an infinite number of eigenvalues and corresponding eigenmodes \cite{HOV05}.


\section{Results}
\label{results}
\subsection{Fixed point stabilization in the nematic phase}
\label{stabilization_nematic}
%
%

\begin{figure}
\includegraphics[width=\columnwidth]{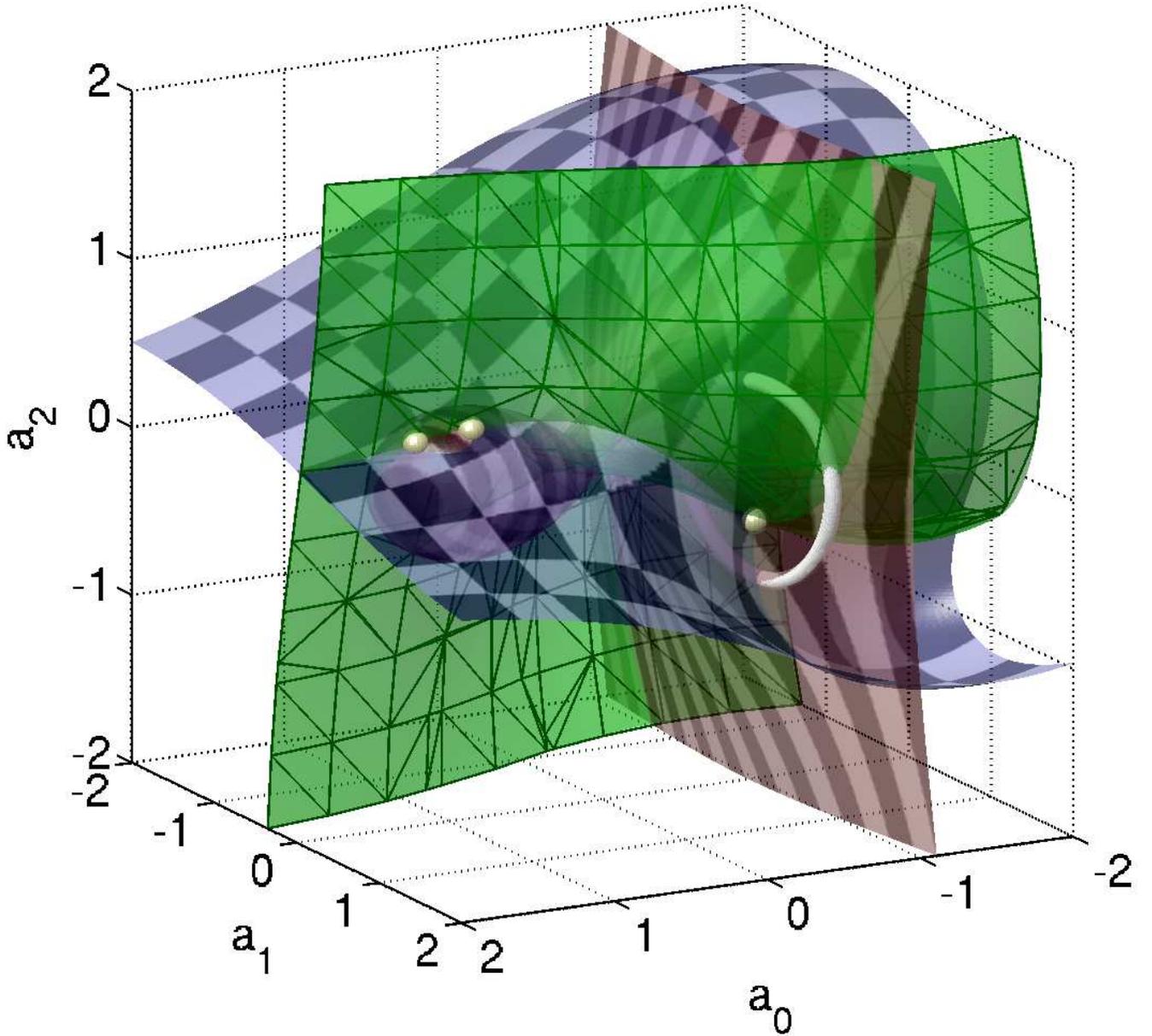}
\caption{\label{fig:Nullclines_theta0}
(Color online) Nullclines corresponding to the dynamical variables $a_0$, $a_1$,  $a_2$. The parameters are fixed
to the set ${\bm{\beta}}_{\mathrm{I}}$ located within the in-plane W regime (see Fig.~\ref{fig:parameterset1}).
The striped, unichrome, and checkerboard patterned surface are obtained from setting $\dot{a_0}= 0$, $\dot{a_1}= 0$, and $\dot{a_2}= 0$, respectively
(the components $a_3$ and $a_4$ vanish at  ${\bm{\beta}}_{\mathrm{I}}$).
The small spheres indicate the intersections of all nullclines and thus depict the (unstable) fixed points of the system,
$\bm{q}^\star_1=(1.30034,0.139206,0.318084,0,0)$, $\bm{q}^\star_2=(-0.563386,0.807425,-0.217522,0,0)$, and $\bm{q}^\star_3=(0.98609,0.232099,0.38358,0,0)$.
The white cycle corresponds to the stable limit cycle emerging around $\bm{q}^\star_2$.}
\end{figure}

%
We start by determining the (unstable) fixed points at the parameter set ${\bm{\beta}}_{\mathrm{I}}$ 
(see Fig.~\ref{fig:parameterset1}).  Since we are in the regime of in-plane dynamic states
($a_3=a_4=0$) we can visualize the nullclines of the system, i.e. the geometrical shapes where $\dot{a}_i=0$,  as two-dimensional surfaces 
in the three-dimensional space spanned by $(a_0,a_1,a_2)$. These surfaces are shown in Fig.~\ref{fig:Nullclines_theta0}.

The fixed points of the system are located where all of the three nullclines intersect. As seen from Fig.~\ref{fig:Nullclines_theta0}, there are three fixed points (indicated by small
spheres). An analysis of the corresponding Jacobian shows that all of these fixed points are unstable (i.e., at least one eigenvalue has a positive real part), as expected in the wagging regime. We remark in this context that the fixed point $\bm{q}^\star_1$ would actually be {\it stable} if 
we had restricted ourselves to the analysis of the 3-dimensional system ($a_0,a_1,a_2$). Indeed, in this case, all three eigenvalues related to $\bm{q}^\star_1$ have negative
real parts. The corresponding "log-rolling" state has been analyzed in Ref.~\cite{Rien02a}. In the full, five-dimensional analysis, however,
this fixed point becomes unstable since the nematic director can "escape" in further directions. The other fixed points, $\bm{q}^\star_2$ and $\bm{q}^\star_3$, are unstable
in both the three- and the five-component dynamical system.

\begin{figure}
\includegraphics[width=3.8cm]{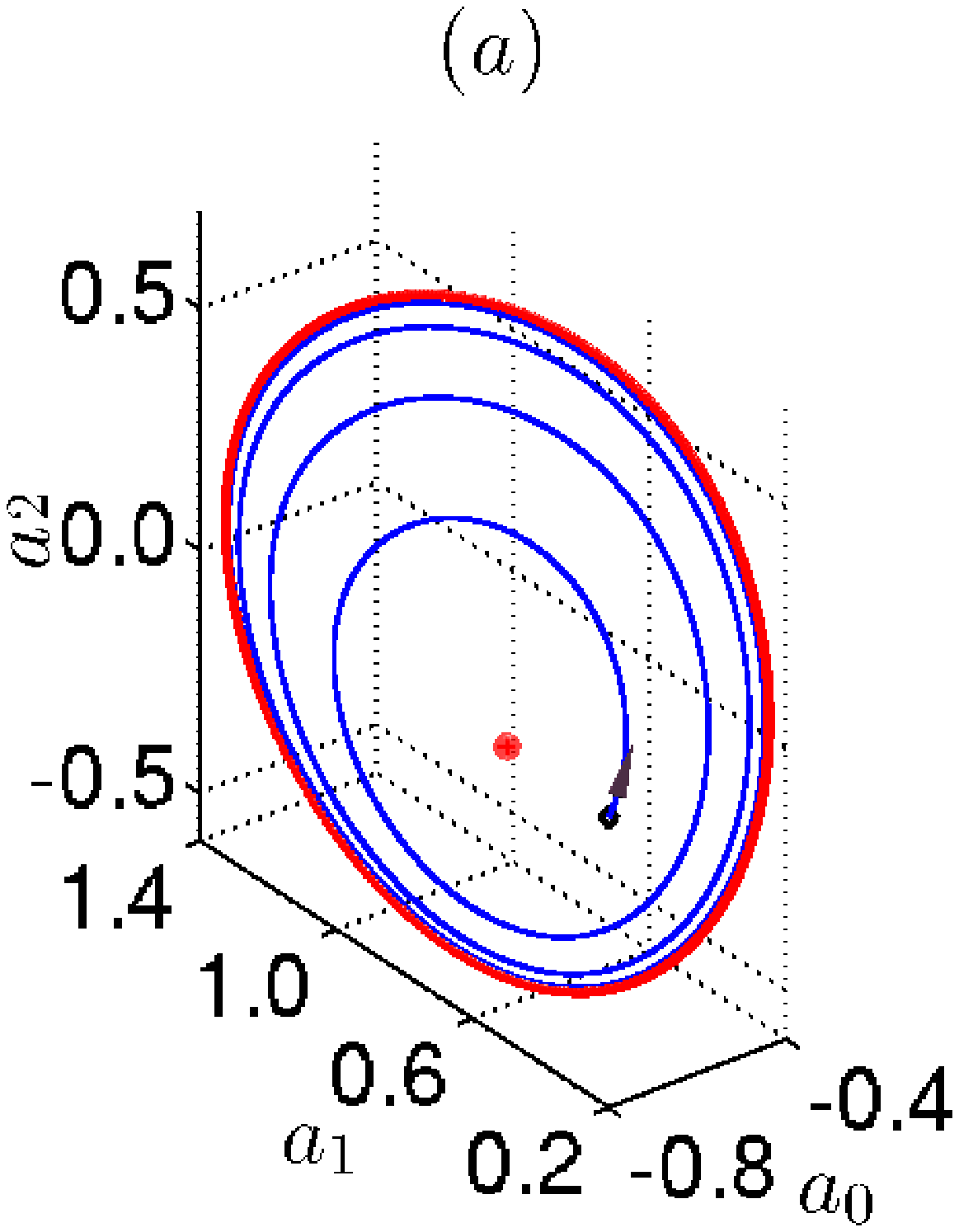}
\includegraphics[width=3.8cm]{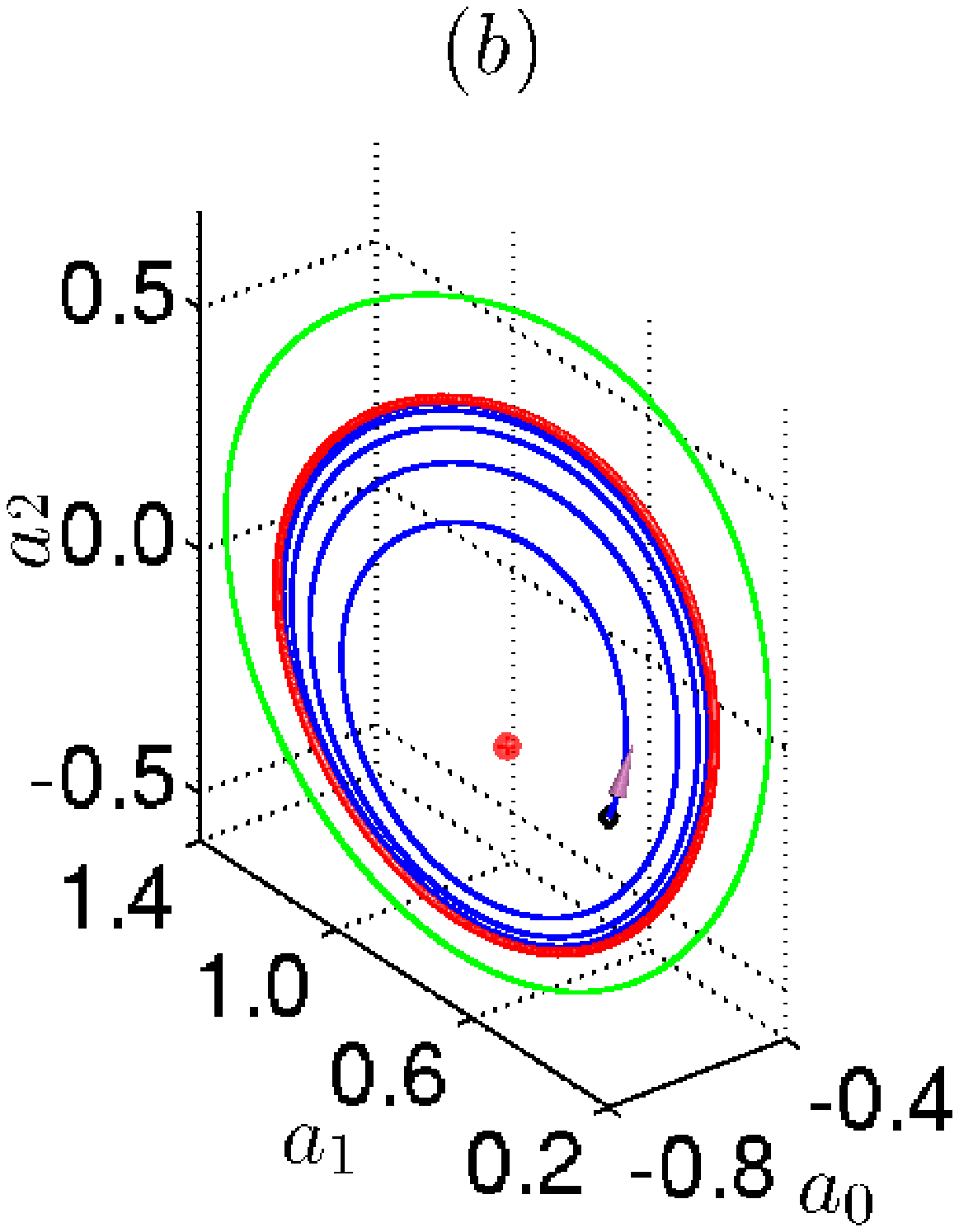}
\includegraphics[width=3.8cm]{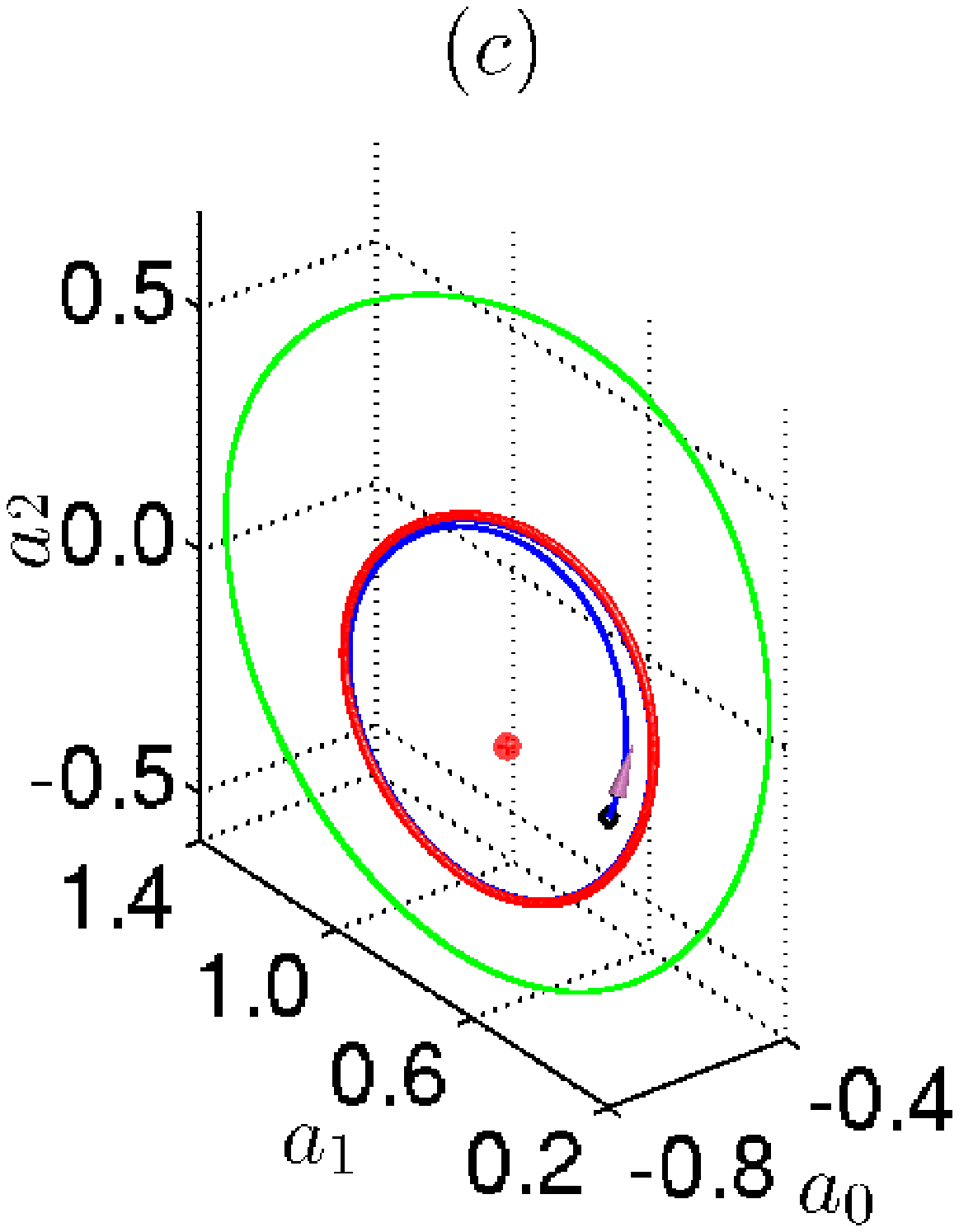}
\includegraphics[width=3.8cm]{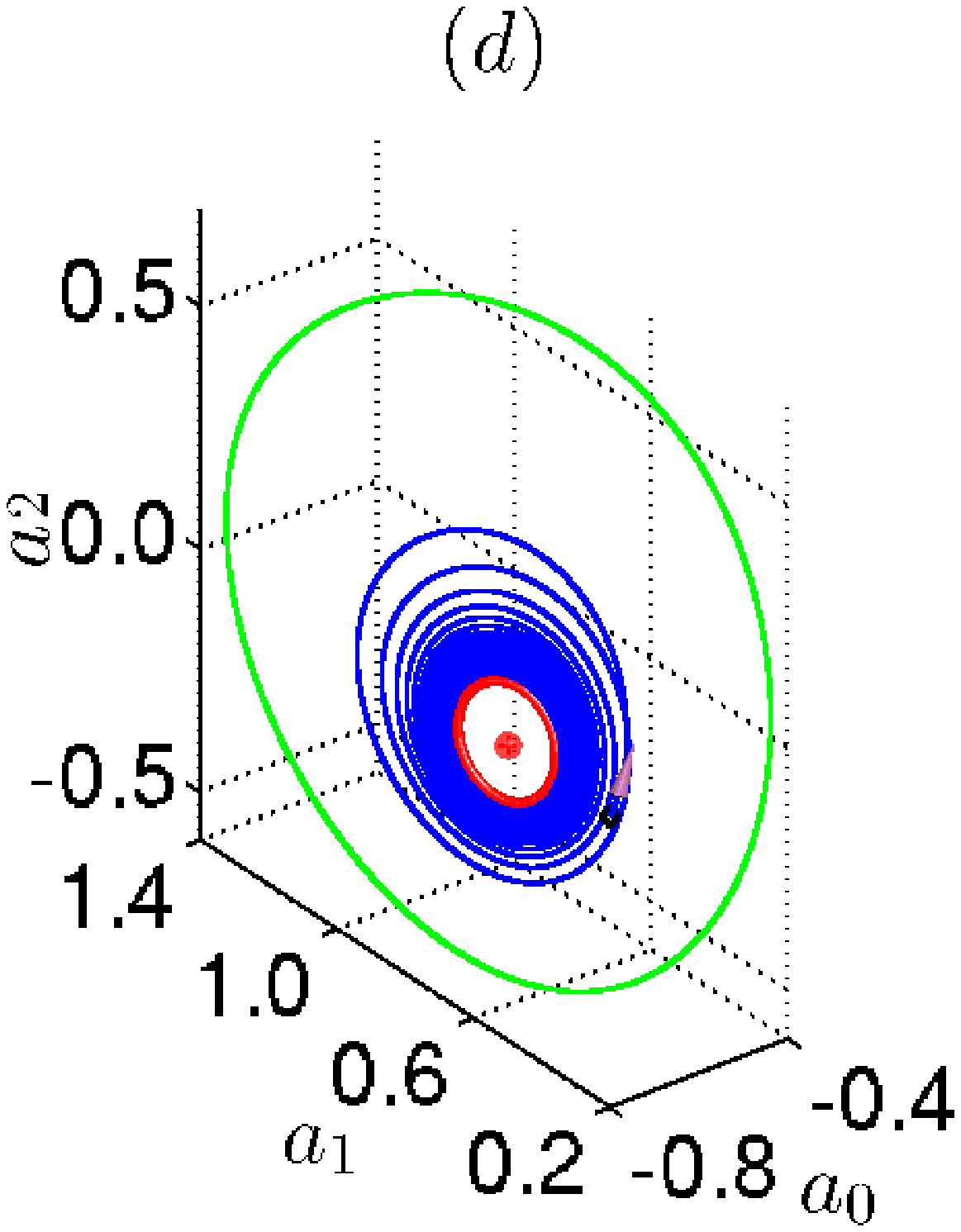}
\includegraphics[width=3.8cm]{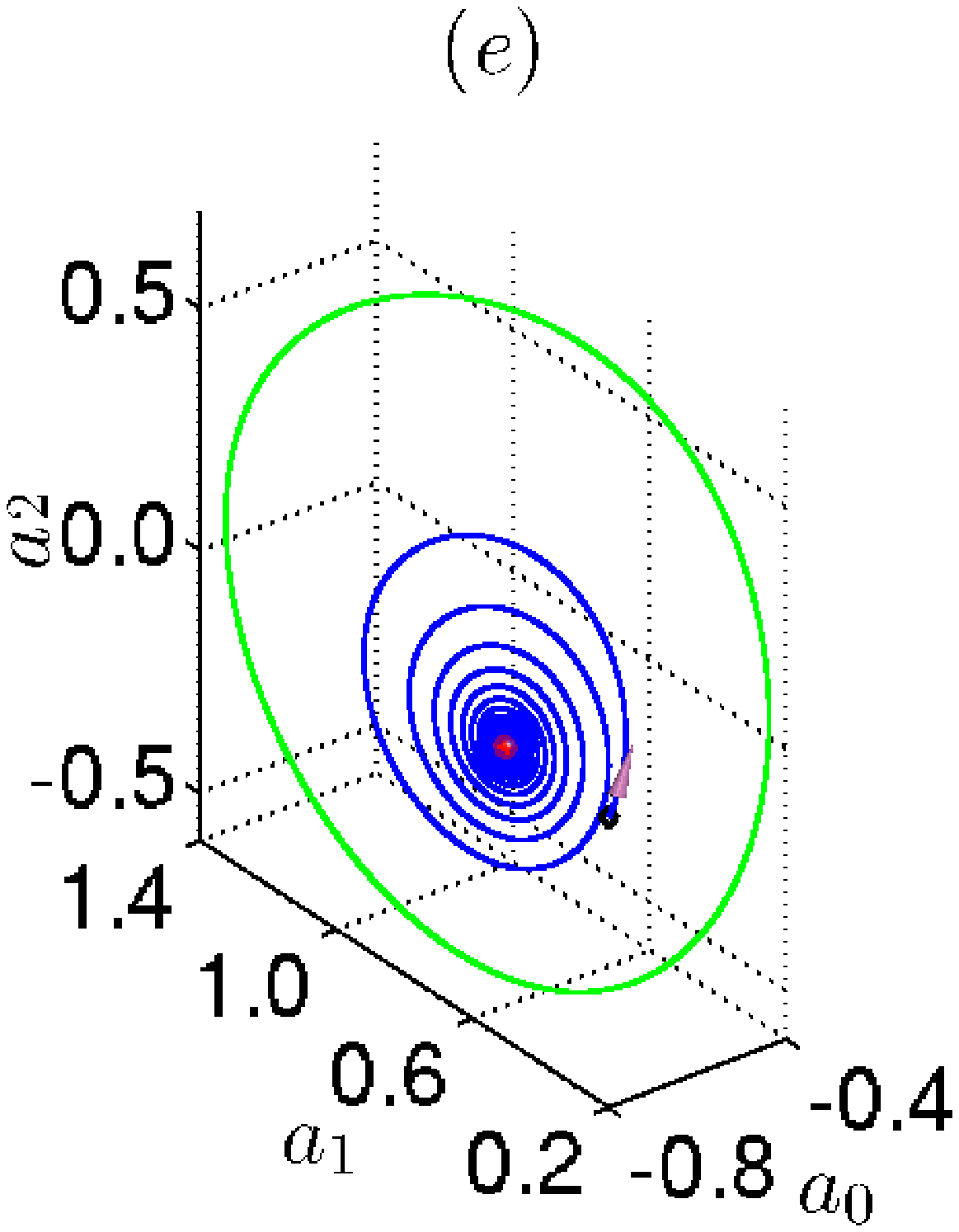}
\includegraphics[width=3.8cm]{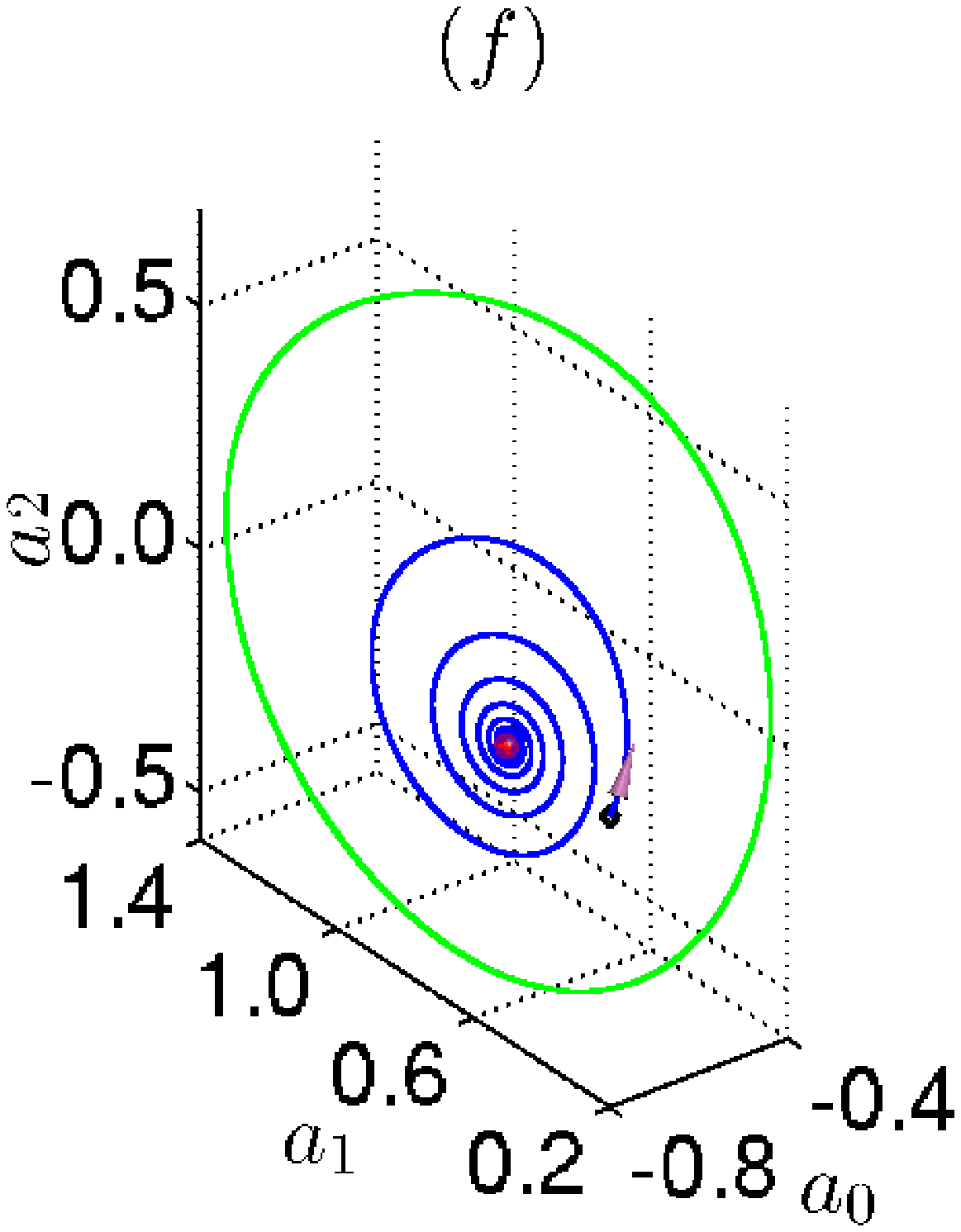}
 \caption{\label{fig:orbits_set1}
  (Color online) Phase portraits of the dynamical variables $a_i$ ($i=0,1,2$). 
  Parameters are fixed to the set ${\bm{\beta}}_{\mathrm{I}}$ located within the in-plane W regime (see Fig.~\ref{fig:parameterset1}).
  The central small (red) dot marks the (unstable) fixed point. (a) Uncontrolled system ($K=0$) with initial condition $a_{0}^{\mathrm{init}}=-0.57$, $a_{1}^{\mathrm{init}}=0.5$, $a_{2}^{\mathrm{init}}=-0.22$.
 The initial condition is indicated with a black diamond. 
(b)-(f) Systems under time-delayed feedback control [see Eq.~(\ref{eq:control})] with the control strength (b) $K=0.1$, (c) $0.2$, (d) $0.3$, (e) $0.4$, and (f) $0.5$, respectively, and delay time $\tau=0.5$. 
In all cases (b)-(f), the control starts at $t=0$, assuming that $a_i(t)=a_i^{\mathrm{init}}$ in the interval $[-\tau,0]$.
In (a)-(d) the blue (dark) trajectories approach stable limit cycles, which are plotted as red (thick dark) cycles. In (b)-(f) the limit cycle of the uncontrolled system (a) is replotted for reference as green (light) cycle. For (e),(f) the trajectories end in the fixed point.}
\end{figure}

Also indicated in Fig.~\ref{fig:Nullclines_theta0} is the (stable) limit cycle emerging around the fixed point $\bm{q}^\star_2$. 
This limit cycle corresponds to undamped oscillations of the order parameters as functions of time. The corresponding period is close to that predicted by linear stability analysis,
$T_0=2\pi/|\mathrm{Im}(\nu)|$, where $\nu$ is one member of the (complex conjugate) pair of eigenvalues at $\bm{q}^\star_2$ that have a positive real part 
(numerically, we find $\nu \approx 0.49+\mathrm{i} 4.61$, yielding $T_0 \approx 1.36$). The dynamical evolution of an unstable configuration of dynamical variables 
towards the limit cycle is illustrated in Fig.~\ref{fig:orbits_set1}~a).
We now apply the TDFC scheme described in Sec.~\ref{delay}. To illustrate the impact on the phase portraits, we show in Figs.~\ref{fig:orbits_set1} b)-f) exemplary results 
for a fixed delay time, $\tau=0.5$, and different values of the control strength, $K$. All calculations have been started with the same initial values for the order parameters,
$a_0$, $a_1$, $a_2$, and the same history regarding the onset of control. Inspecting Figs.~\ref{fig:orbits_set1} it is seen that for $K\leq 0.3$, the feedback control reduces the 
diameter of the limit cycle but the dynamics remains oscillatory at long times [see parts (b)-(d)]. However, for $K=0.4$ and $K=0.5$, the initially oscillatory motion becomes 
more and more damped out with time, and the final state is the fixed point $\bm{q}^\star_2$. Physically, this means that the director freezes along an in-plane direction, corresponding 
to flow alignment. Thus, the control scheme has been successful. 

A more systematic way to analyze the stability of the fixed point $\bm{q}^\star_2$ under feedback control is to monitor the complex eigenvalues $\mu$
[see Eq.~(\ref{eigen_control})]. Specifically, we consider
the largest real part of $\mu$. Indeed, due to the transcendental character of Eq.~(\ref{eigen_equation}), the spectrum of eigenvalues of the controlled system is infinite due to the infinite number of branches of the Lambert ${\cal W}$ function (multileaf structure). Stabilization (within the linear approximation) of the fixed point 
then means that $\mathrm{max}\left(\mathrm{Re}\left(\mu\right)\right)$ is negative at the values of $K$ and $\tau$ considered.

\begin{figure}
\includegraphics[width=\columnwidth]{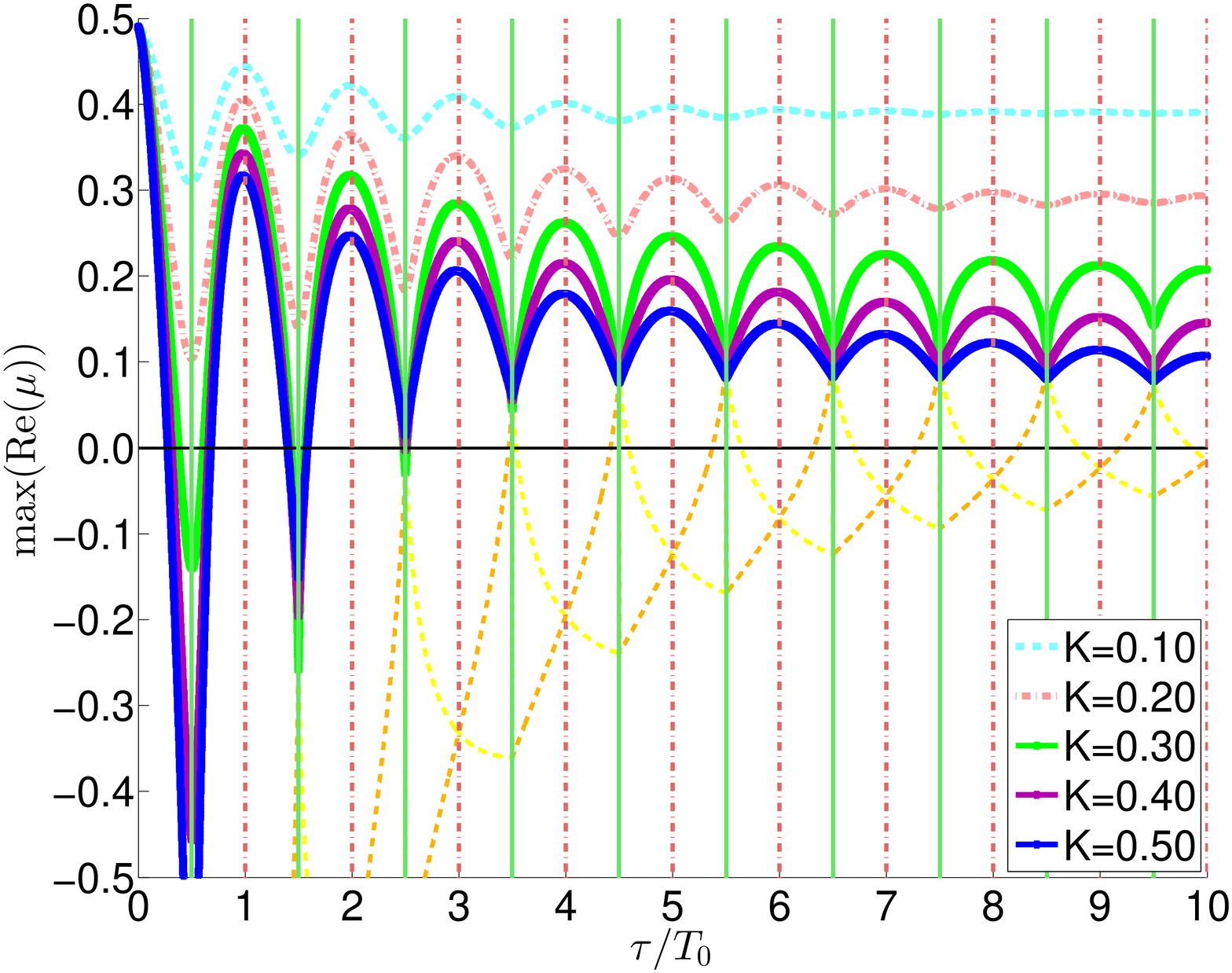}
\caption{\label{fig:Ktau_Eigenvalue_theta0}
(Color online) Largest real part of the complex eigenvalues $\mu$ vs $\tau$ for different values of $K$. The dot-dashed vertical lines correspond to multiples of $T_0=2\pi/|\mathrm{Im}(\nu)|$ ($T_0 \approx 1.36$), where $\nu$ is the eigenvalue of the uncontrolled system at the fixed point $\bm{q}^\star_2$ (see Fig.~\ref{fig:Nullclines_theta0}). 
The solid vertical lines are shifted relative to the dashed lines by $T_0/2$.
The dashed lines in the lower part indicate some lower branches of eigenvalues for $K=0.30$.
The other parameters are fixed to the set ${\bm{\beta}}_{\mathrm{I}}$ located within the in-plane W regime (see Fig.~\ref{fig:parameterset1}).}
\end{figure}

To get a first insight into the ranges of control parameters, where TDFC works, we plot in Fig.~\ref{fig:Ktau_Eigenvalue_theta0} the quantity
$\mathrm{max}\left(\mathrm{Re}\left(\mu\right)\right)$ as a function of $\tau$ for different values of the control strength $K$. 
All curves start at the value of the
real part corresponding to $\tau = 0$ where the control
terms in Eq.~(\ref{eq:control}) vanish. The positive value 
indicates the instability of the steady state. Upon increase
of the delay time from zero, the largest real parts corresponding to different $K$ first
decrease up to a certain delay time, and display subsequently
an oscillatory behavior. However, for the cases $K=0.1$ and $K=0.2$, the largest real part remains positive. Only for $K\geq 0.3$ 
there exist values of $\tau$ where the largest
real parts become negative, indicating that the control is successful. 

Following the analysis in Ref.~\cite{HOV05}, it is possible to analytically determine the minimum value of $K$ required to control the system at specific values of $\tau$. 
The boundary of stability is determined by the condition $\mathrm{Re}(\mu)=0$, or equivalently, 
$\mu=i\Omega$ (with $\Omega$ being real). Inserting this into Eq.~(\ref{eigen_equation})
we find 
\begin{eqnarray}
\label{boundary}
\mathrm{Re}(\nu) & = & K\left(1-\cos\Omega\tau\right)\nonumber\\
\mathrm{Im}(\nu) & = & \Omega+K\sin\Omega\tau.
\end{eqnarray}
From the first equation of Eq.~(\ref{boundary}) it follows that at the boundary of stability, $K$ varies between $\mathrm{Re}(\nu)/2$ and $\infty$
(since $\cos\Omega\tau$ is bounded between $-1$ and $1$). Thus, the minimal value of $K$ is given by 
\begin{equation}
\label{Kmin}
K_{\mathrm{min}}=\mathrm{Re}(\nu)/2
\end{equation}
corresponding to $\cos\Omega\tau=-1$. The fixed point $\bm{q}^\star_2$ considered here is characterized by $\mathrm{Re}(\nu)\approx 0.49$, yielding
$K_{\mathrm{min}}\approx 0.25$ according to Eq.~(\ref{Kmin}). This is consistent with the results displayed
in Fig.~\ref{fig:Ktau_Eigenvalue_theta0}.
We can also obtain a constraint for the delay times corresponding to the stability boundary. Specifically, the condition $\cos\Omega\tau=-1$ requires that
\begin{equation}
\label{omega}
\Omega\tau=(2n+1)\pi,\quad n=0,1,2,\ldots
\end{equation}
Equation~(\ref{omega}) immediately implies that $\sin\Omega\tau=0$. From the second equation of Eq.~(\ref{boundary}) it therefore follows
that $\Omega=\mathrm{Im}(\nu)$. Combining this with Eq.~(\ref{omega}) we obtain the following condition for the delay times at minimum K:
\begin{eqnarray}
\label{tau}
\tau&=&\frac{\pi}{\mathrm{Im}(\nu)}(2n+1)\nonumber\\
& = & T_0\left(n+\frac{1}{2}\right),\quad n=0,1,2,\ldots,
\end{eqnarray}
where we have used that $T_0=2\pi/|\mathrm{Im}(\nu)|$. We conclude that both of the control parameters, $K$ and $\tau$, required to stabilize the fixed point are intimately related to the
eigenvalues of the uncontrolled system

A further interesting case occurs when $\cos\Omega\tau=1$ in Eq.~(\ref{boundary}). In this case, control is essentially impossible, since the corresponding value of the control strength [at finite values of $\mathrm{Re}(\nu)$] is $K=\infty$. The corresponding delay times can be found using the same arguments as those leading to Eq.~(\ref{tau}). Specifically, one
has $\sin\Omega\tau=0$ (and thus, $\Omega=\mathrm{Im}(\nu)=2\pi/T_0$), but this time 
$\Omega\tau$ is an even multiple of $\pi$ [contrary to Eq.~(\ref{omega})]. We therefore find that at delay times
$\tau=n T_0$ with $n=0,1,2,\ldots$, stabilization via TDFC is not possible for any finite value of $K$.
The important role of the delay time is reflected in Fig.~\ref{fig:Ktau_Eigenvalue_theta0}. 
For all values of $K$ considered, we find the minima of the
functions $\mathrm{max(Re(}\mu)$ to occur at $\tau=(n+1/2) T_0$, while maxima occur at even multiples of $T_0$.
Further (analytic) results on the domain of control in the ($K,\tau$)-plane can be found in Ref.~\cite{HOV05}, 
where the same, diagonal control scheme has been employed to stabilize a fixed point.

\begin{figure}
\includegraphics[width=\columnwidth]{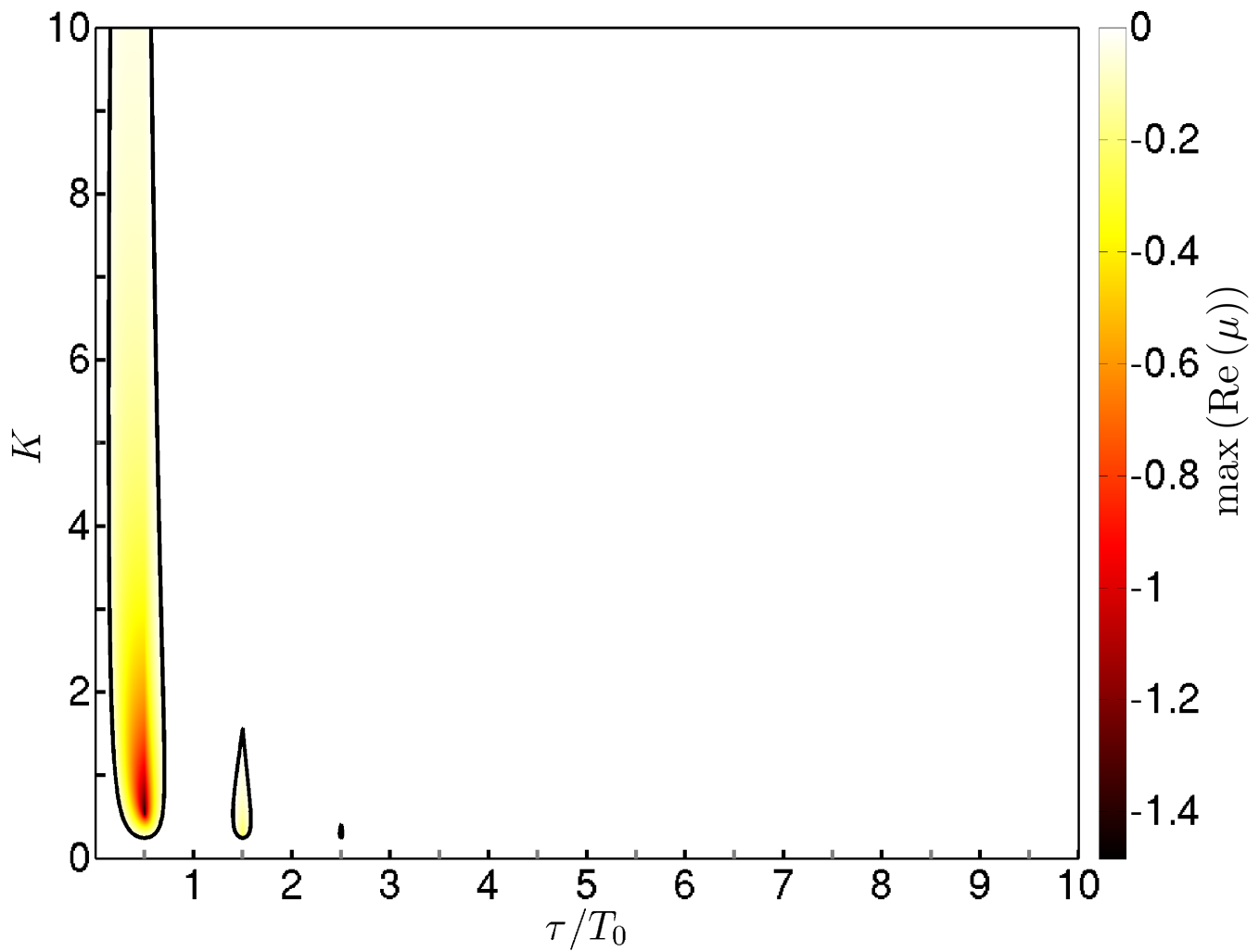}
\caption{\label{fig:Ktau_plane_theta0}
(Color online) Largest real part of the complex eigenvalues $\mu$ in the $K$-$\tau$ plane (only negative values are shown).
The black contours bound regions where the largest real part is negative.
The shear parameters are fixed to the set ${\bm{\beta}}_{\mathrm{I}}$ located within the in-plane W regime (see Fig.~\ref{fig:parameterset1}).}
\end{figure}

So far we have focused on some specific values of $K$. To get an overall "stability map" we show in Fig.~\ref{fig:Ktau_plane_theta0} the real part of the largest eigenvalue 
in the $K-\tau$ plane (only negative values are plotted).
The black contours indicate the control parameters where the real part of the largest eigenvalue becomes zero.
Within the black contours the real part is negative (colored in the plot); therefore these
lines bound the regions where TDFC is successful. It is seen that the areas of stabilization shrink with increasing delay time and eventually disappear.
This is due to the scaling behavior of the eigenvalue spectrum for large $\tau$ \cite{YAN06,WOL10}.
\subsection{Stabilization in the isotropic phase}
\begin{figure}
 \includegraphics[width=\columnwidth]{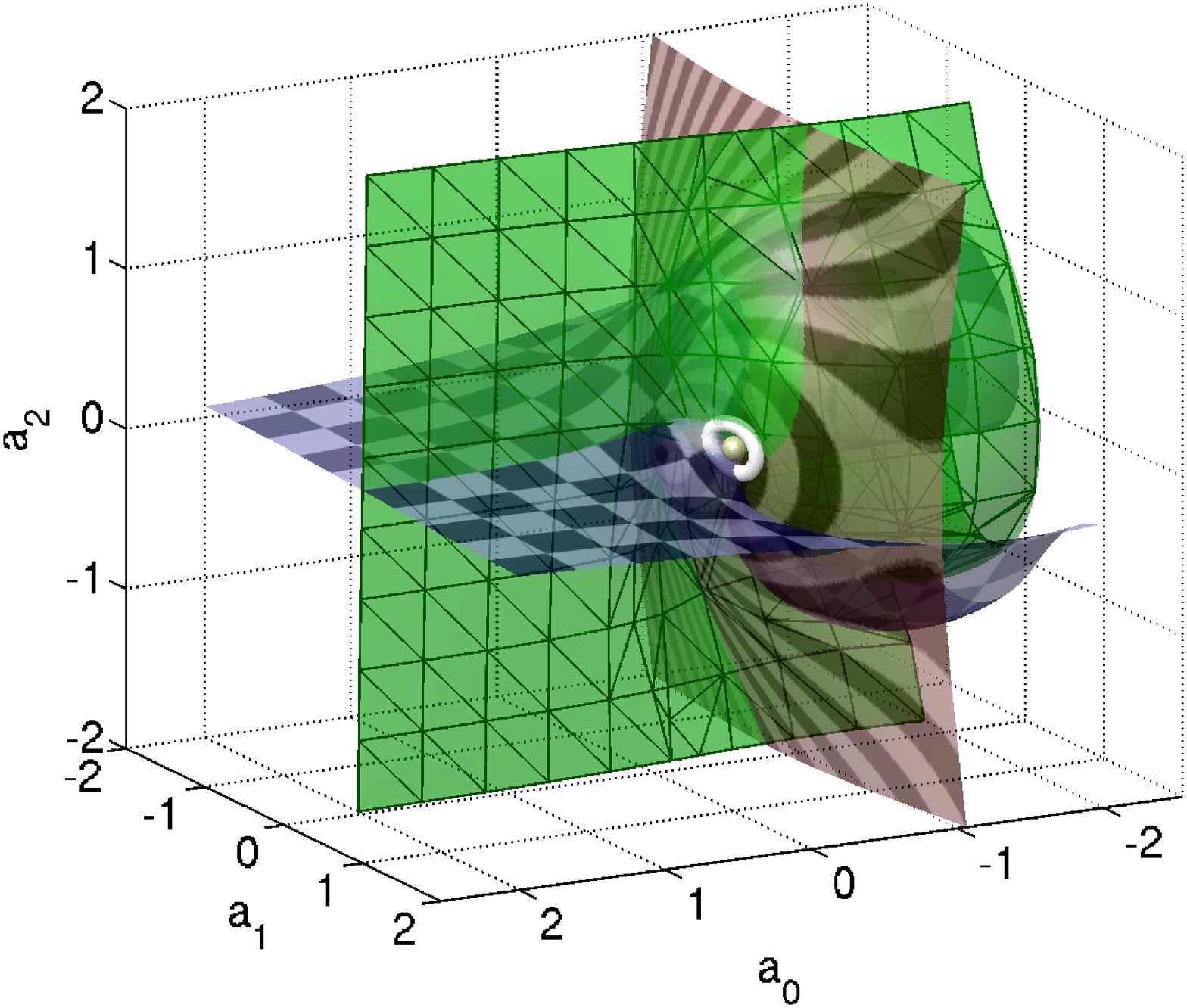}
\caption{\label{fig:Nullclines_theta120}
(Color online) Same as Fig.~\ref{fig:Nullclines_theta0}, but for the parameter set
${\bm{\beta}}_{\mathrm{II}}$ located within the island of W dynamics, which appears on shearing from the isotropic phase 
(see Fig.~\ref{fig:parameterset2}). The small sphere indicates the unstable fixed point,
$\bm{q^*}_1=[-0.283501,0.49587,0.108511,0,0]$.
The white cycle corresponds to the stable limit cycle emerging around $\bm{q}^\star_1$.}
\end{figure}

As a second example for the stabilization of a fixed point we now consider the reduced temperature $\theta=1.20$, at which the 
equilibrium (i.e., zero-shear) system is orientationally disordered. The presence of shear then induces either flow alignment or oscillatory dynamics (of type W). As seen
from Fig.~\ref{fig:parameterset2}, stable W motion occurs at rather small values of the coupling parameter $\lambda_{\mathrm{K}}$ and intermediate values
of the shear rate. Within this "island" of W motion, we now focus on the parameter set ${\bm{\beta}}_{\mathrm{II}}$. The corresponding nullclines  
(pertaining to the components $a_0$, $a_1$, and $a_2$) are shown in Fig.~\ref{fig:Nullclines_theta120}. 
Contrary to the situation within the nematic phase
discussed in the previous paragraph, we find at ${\bm{\beta}}_{\mathrm{II}}$
only one (unstable) fixed point,  $\bm{q^*}_1$, and one stable limit cycle. 
The relevant eigenvalue of the uncontrolled system at the fixed point is given by $\nu \approx 0.027+i0.65$.
From that, we find the oscillation period $T_0=2\pi/|\mathrm{Im}(\nu)|\approx 9.68$.

\begin{figure}
\includegraphics[width=4cm]{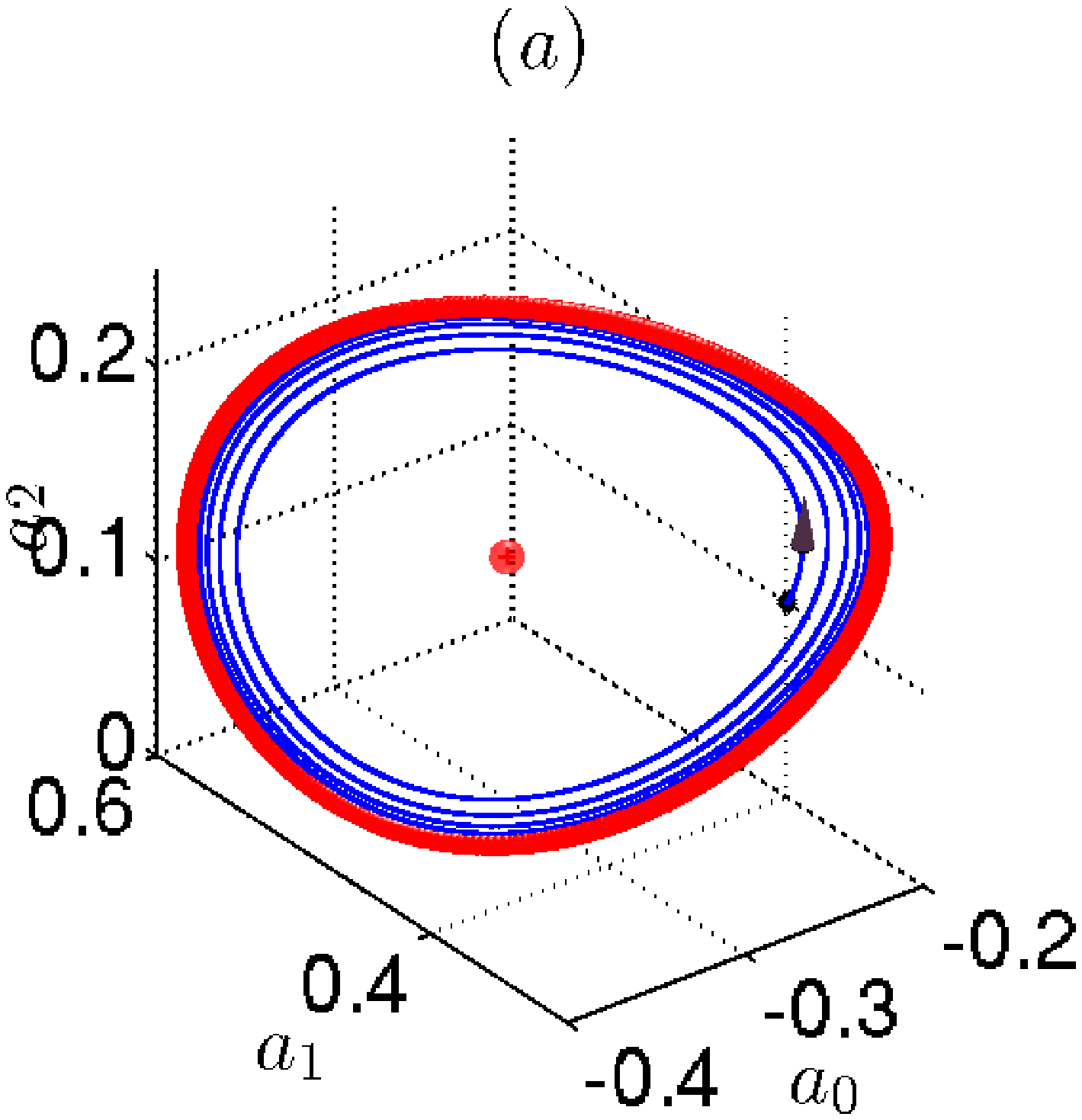}
\includegraphics[width=4cm]{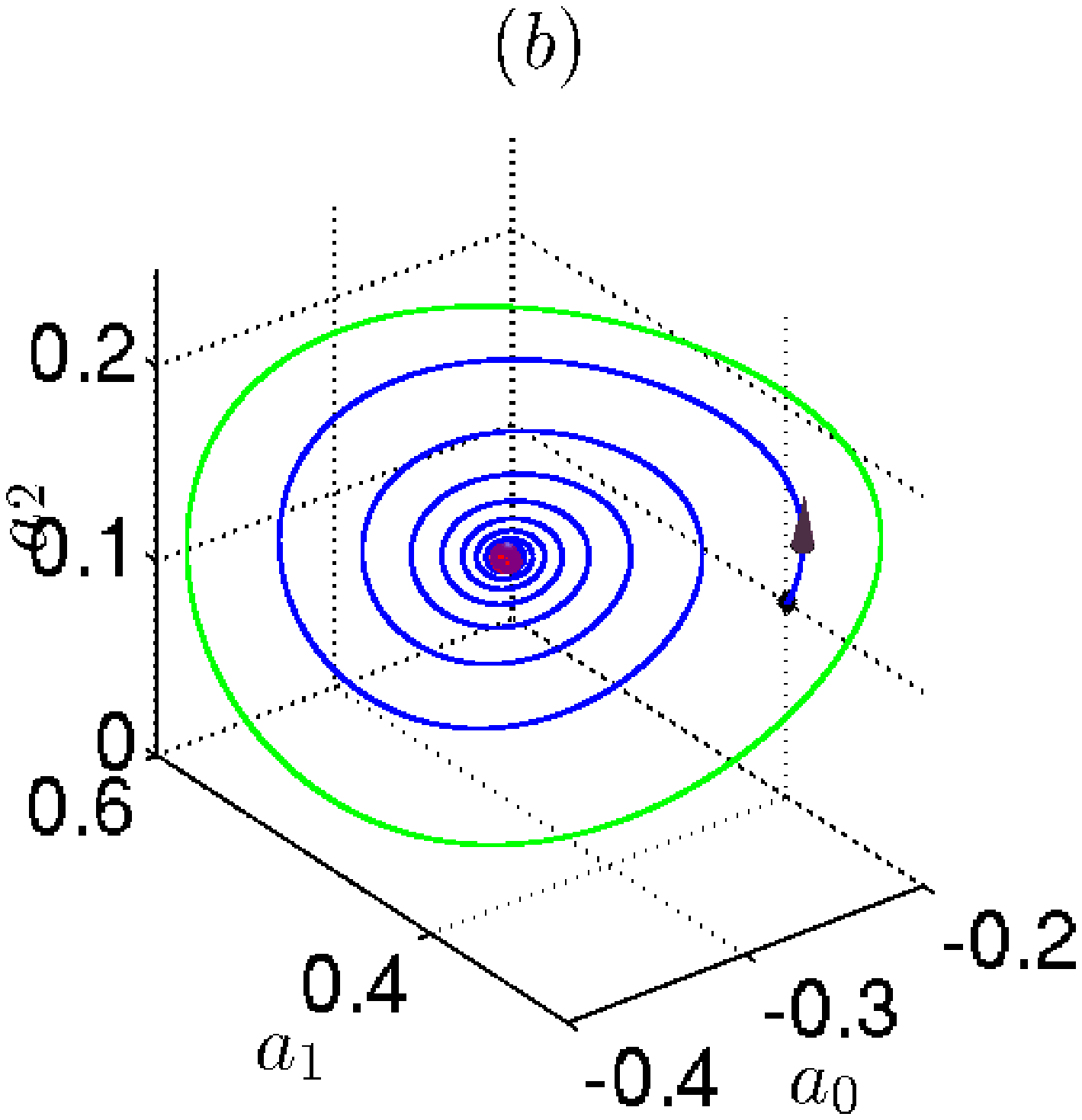}
\includegraphics[width=4cm]{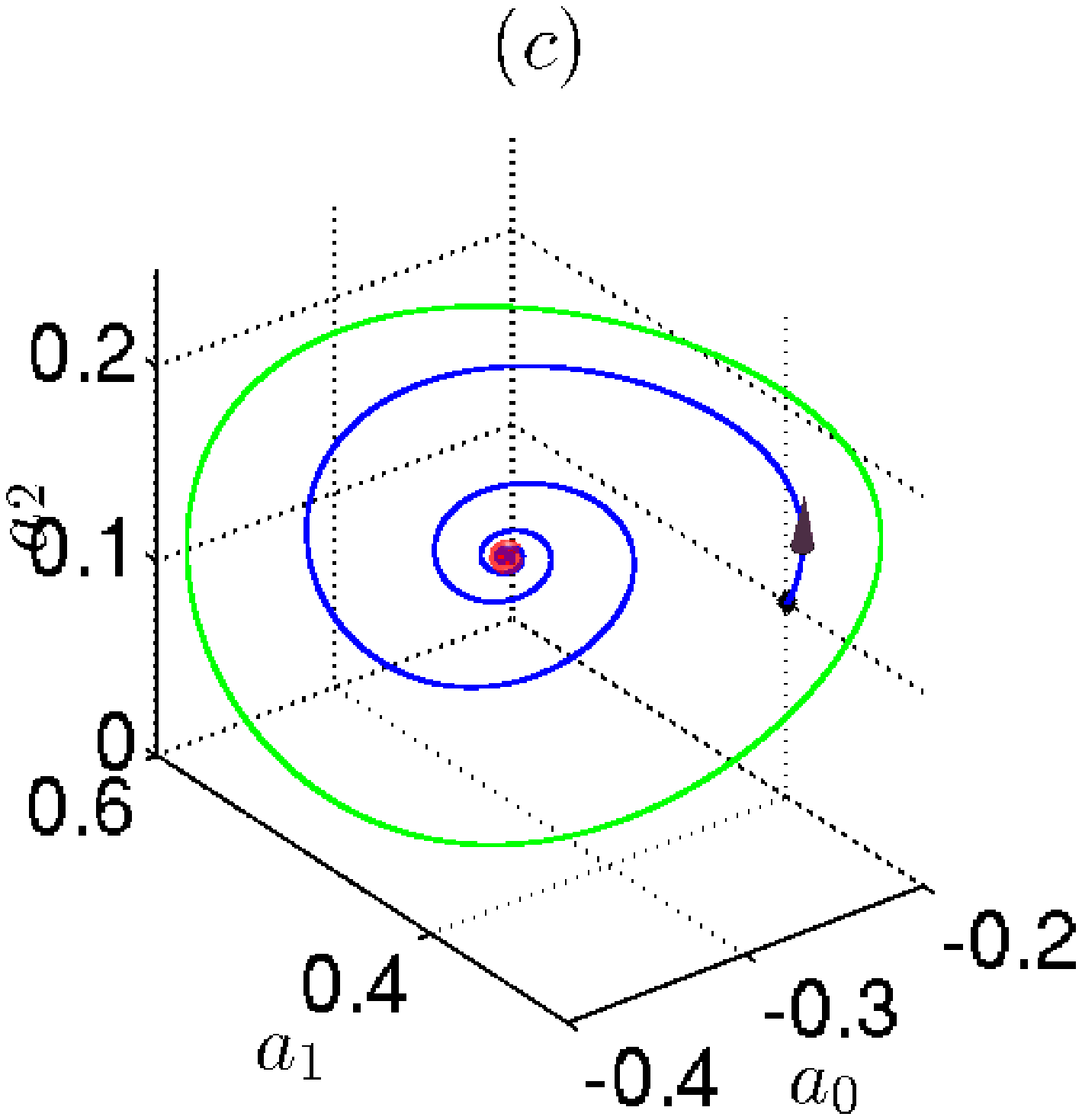}
\includegraphics[width=4cm]{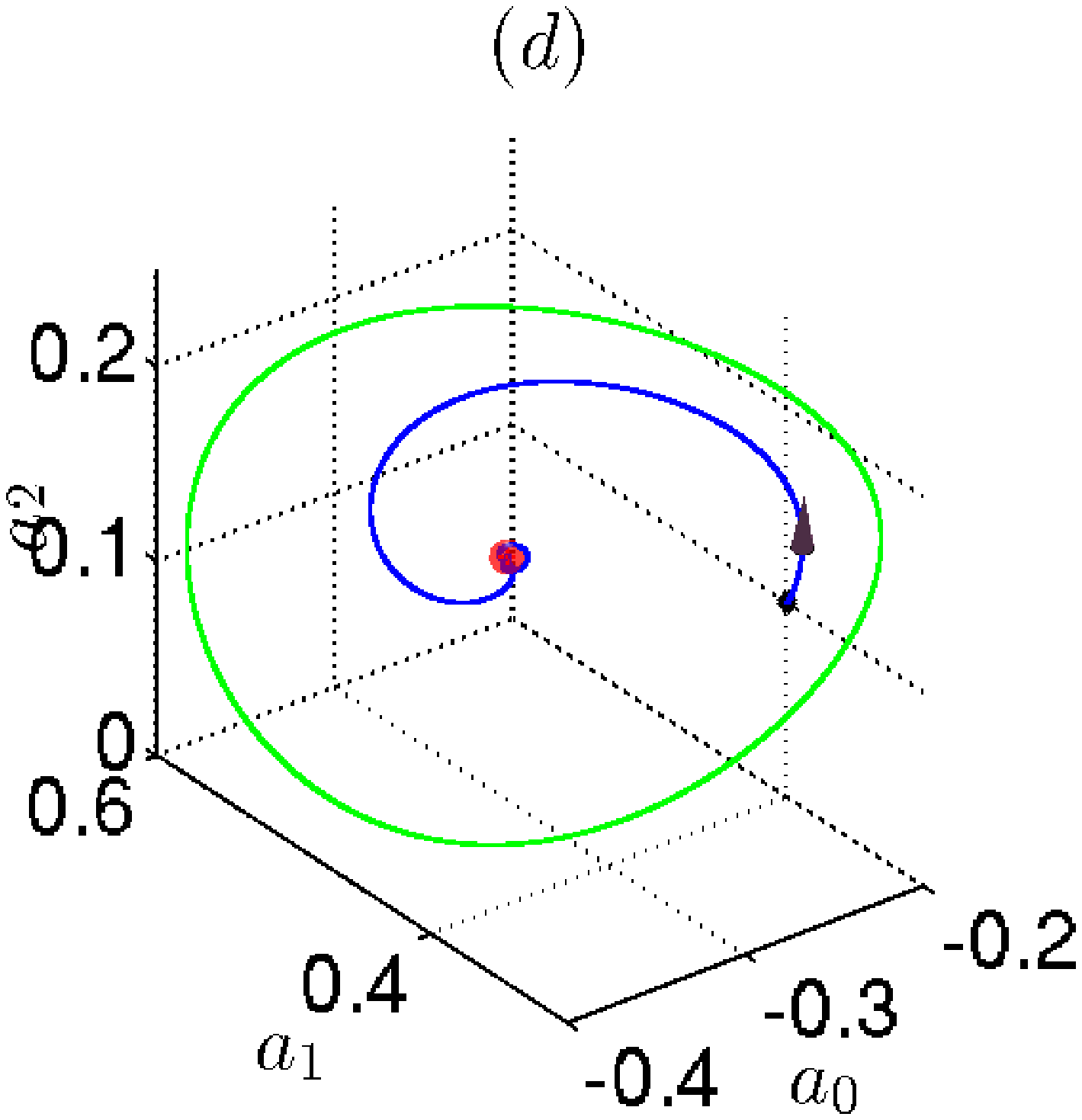}
\includegraphics[width=4cm]{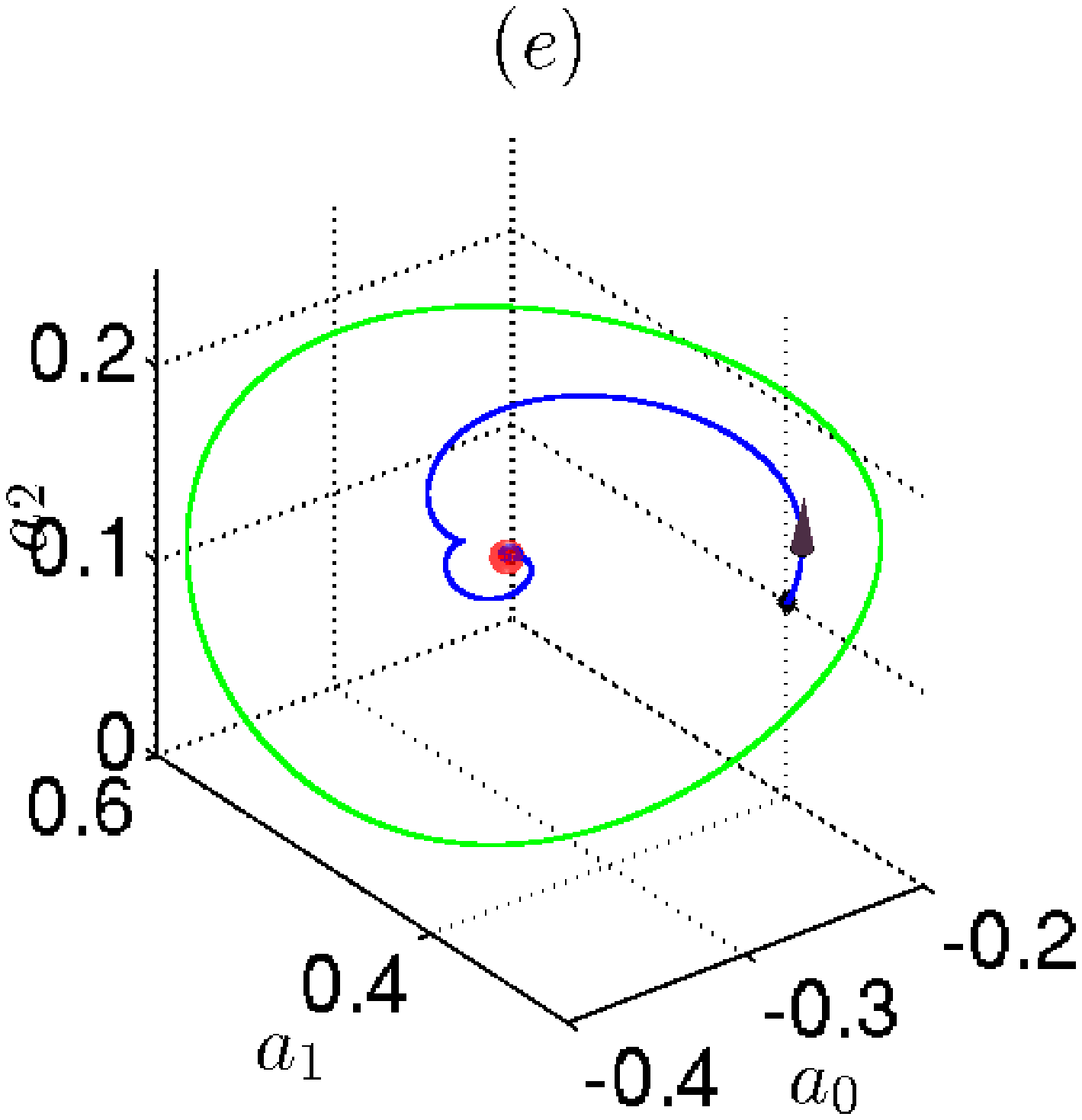}
\includegraphics[width=4cm]{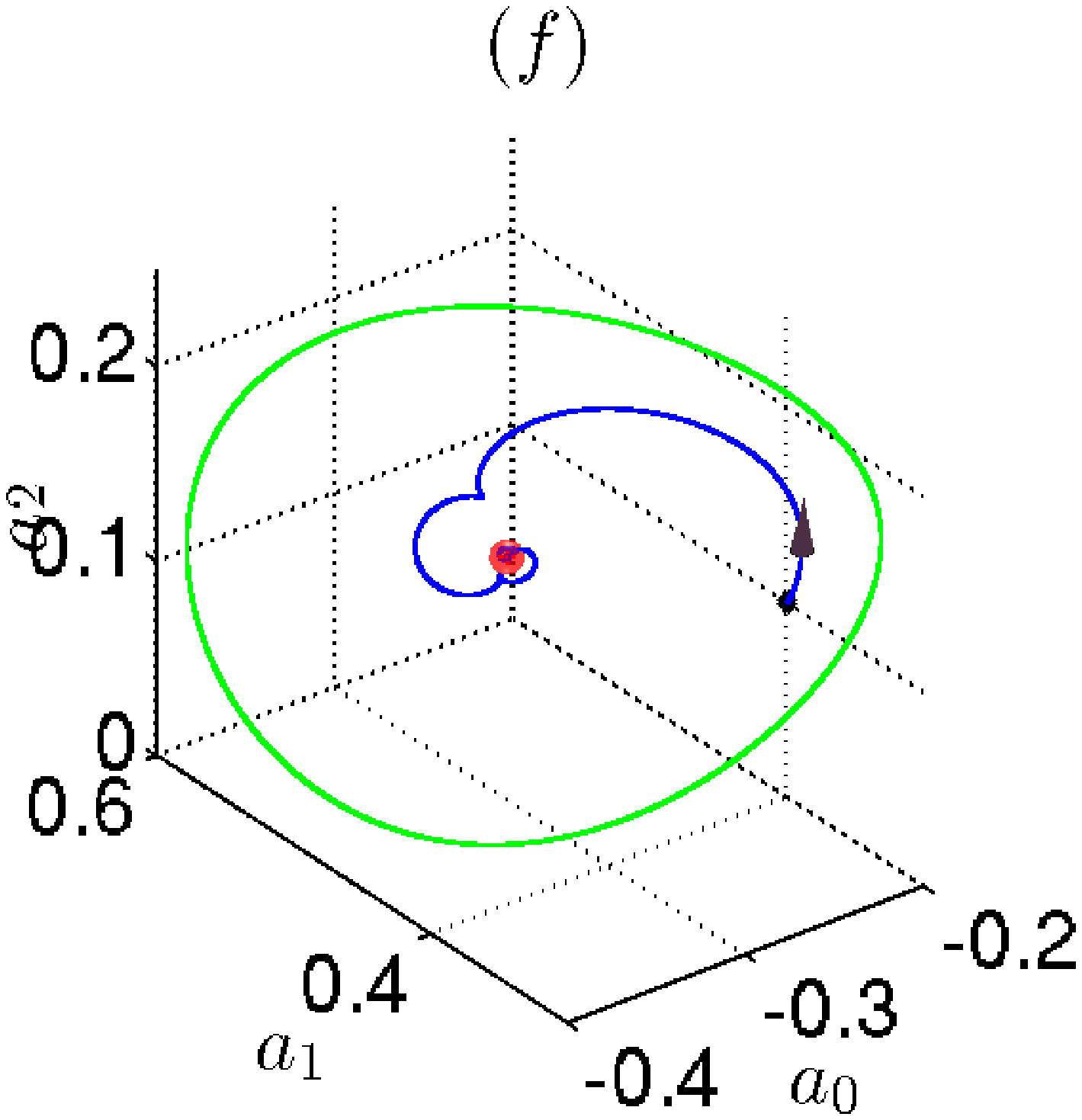}
 \caption{\label{fig:orbits_set2}
 (Color online) Phase portraits of the dynamical variables $a_i$ ($i=0,1,2$). The central small dot marks the (unstable) fixed point. (a) Uncontrolled system ($K=0$) with initial condition $a_{0}^{\mathrm{init}}=-0.20$, $a_{1}^{\mathrm{init}}=0.4$, $a_{2}^{\mathrm{init}}=0.1$. 
(b)-(f) Systems under time-delayed feedback control [see Eq.~(\ref{eq:control})] with the control strength (b) $K=0.03$, (c) $0.05$, (d) $0.1$, (e) $0.15$, and (f) $0.2$, respectively, and delay time $\tau=5.0$. 
In all cases (b)-(f), the control starts at $t=0$, assuming that $a_i(t)=a_i^{\mathrm{init}}$ in the interval $[-\tau,0]$. In (a) the blue (dark) trajectory approaches a stable limit cycle, that is plotted as red (thick dark) cycle.
In (b)-(f) the limit cycle of the uncontrolled system (a) is replotted for reference as green (light) cycle.
For (b)-(f) all trajectories end in the fixed point.
The shear parameters are fixed to the set ${\bm{\beta}}_{\mathrm{II}}$, see Fig.~\ref{fig:parameterset2}.}
\end{figure}

In Fig.~\ref{fig:orbits_set2}a) we replot this limit cycle, supplemented by phase portraits illustrating the impact of TDFC.
It is seen that the feedback control has a significant effect already at very small values of the control strength, that is, at $K=0.03$, although the chosen time delay is rather
large ($\tau=5.0$). This already indicates that the system reacts more sensitively to TDFC compared to the system considered before.

\begin{figure}
 \includegraphics[width=\columnwidth]{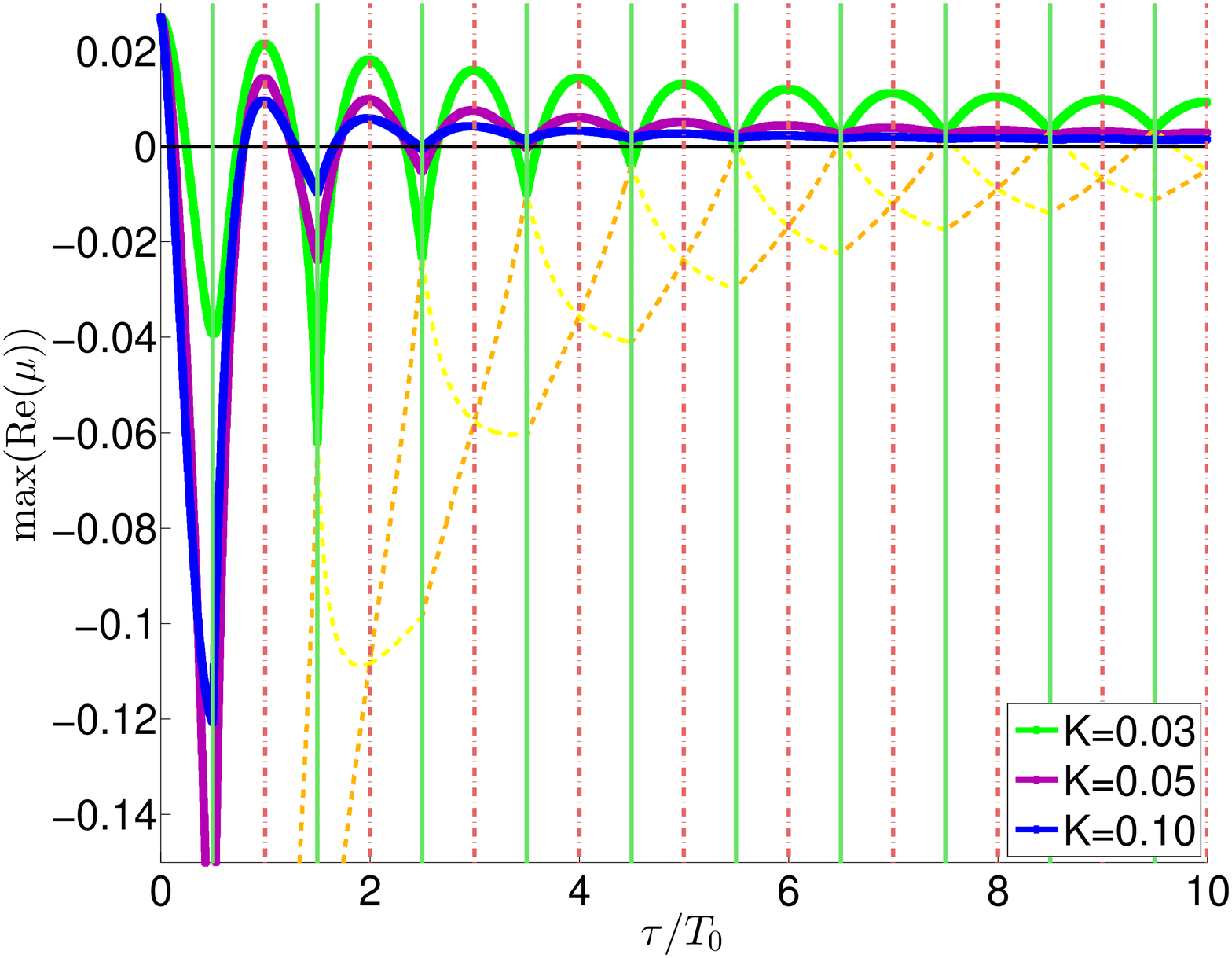}
\caption{\label{fig:Ktau_Eigenvalue_theta120}
(Color online) 
Largest real part of the complex eigenvalues $\mu$ vs $\tau$ for different values of $K$
(the shear parameters are fixed to the set ${\bm{\beta}}_{\mathrm{II}}$, see Fig.~\ref{fig:parameterset2}). The dash-dotted vertical lines correspond to multiples of $T_0=2\pi/|\mathrm{Im}(\nu)|$ ($T_0 \approx 9.68$), where $\nu$ is the eigenvalue of the uncontrolled system at the fixed point $\bm{q}^\star_1$ (see Fig.~\ref{fig:Nullclines_theta120}). 
The solid vertical lines are shifted relative to the dash-dotted lines by $T_0/2$.
The dashed lines in the lower part indicate some lower branches of eigenvalues for $K=0.03$.}
\end{figure}

The fact that small $K$ are sufficient to stabilize the fixed point is supported by the behavior of the largest real part of the eigenvalue $\mu$ 
plotted as function of $\tau$ in Fig.~\ref{fig:Ktau_Eigenvalue_theta120}.
Clearly, the functions have the same qualitative behavior as those obtained in the nematic phase (see Fig.~\ref{fig:Ktau_Eigenvalue_theta0}); 
however, the numerical values of $K$ are much smaller.

\begin{figure}
 \includegraphics[width=\columnwidth]{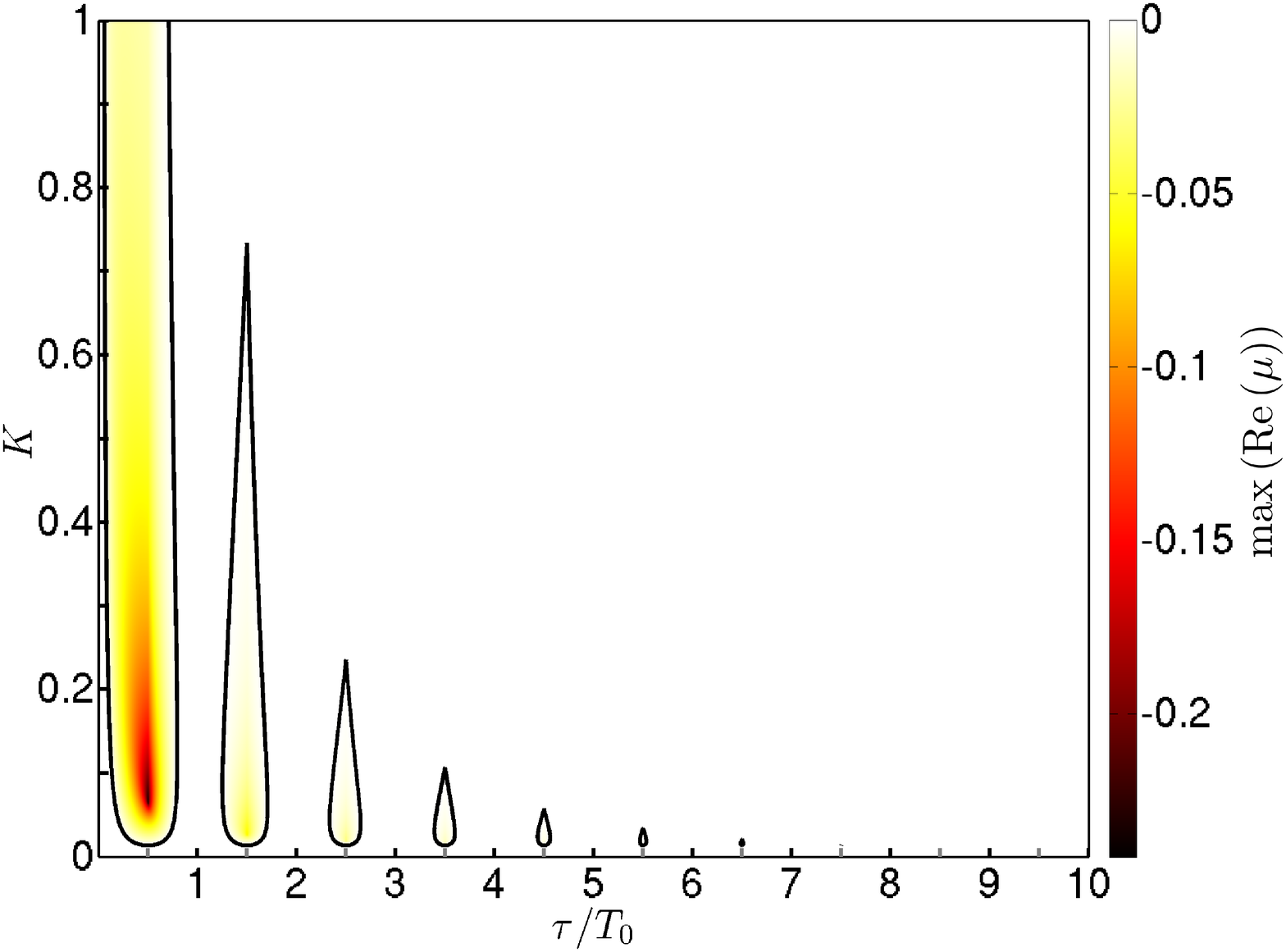}
\caption{\label{fig:Ktau_plane_theta120}
(Color online) Largest real part of the complex eigenvalues $\mu$ in the $K$-$\tau$ plane.
The black contours bound regions where the largest real part is negative, these regions are colored, according to the colormap.
The shear parameters are fixed to the set ${\bm{\beta}}_{\mathrm{II}}$, see Fig.~\ref{fig:parameterset2}.}
\end{figure}

Finally, we present in Fig.~\ref{fig:Ktau_plane_theta120} the ranges of control parameters where TDFC is successful; the boundaries have again been obtained 
from Eqs.~(\ref{Kmin}) and (\ref{tau}), respectively. Compared to the nematic system, we observe a much larger number of areas where the fixed point can be stabilized.
This underlines our finding that less effort is required to stabilize liquid-crystalline systems sheared from the isotropic phase, than systems sheared from the nematic phase.

\section{Conclusions}
\label{conclude}
In the present paper we have discussed a five-dimensional dynamical system describing the director dynamics of sheared liquid crystal under time-delayed feedback control. 
The goal 
was to stabilize the stationary, "flow-aligned" state for shear rates and shear coupling parameters, where the uncontrolled system performs oscillatory wagging motion
in the shear plane. To this end we have applied a diagonal feedback control scheme involving a single delay time $\tau$.  
Following earlier theoretical work \cite{HOV05}, we have analytically studied the (linear) stability problem of the controlled system, yielding explicit expressions 
for the domain of control. One main result is that the {\it optimal} values of 
control strength and delay time are closely related to the relevant complex eigenvalue of the corresponding uncontrolled system, that is,
the sheared liquid crystal in the "wagging" state. 
Interestingly, there is also a strong influence of the equilibrium state from which the
system was sheared: in a system sheared from the isotropic phase, the control strength required to stabilize flow alignment is much smaller than in a sheared nematic system.
This shows that interaction effects between the particles (which are responsible for the isotropic-nematic transition) are crucially important.

From a more general perspective, the present study thus shows that TDFC of unstable steady states, which has already been applied to a variety of optical and neural systems
\cite{DAH10,Lehnert2011,SCH08,HOEV10},
is also possible and useful in the context of non-equilibrium soft-matter systems such as sheared complex fluids. In that case, stabilization of the flow-aligned state seems particularly interesting because flow-alignment is characterized by a small shear viscosity (as opposed to that of oscillatory states). Nevertheless, from a fundamental point of view 
it would be very interesting to extend the present analysis to the (time-delayed) feedback control
of oscillatory states, which are also observed experimentally (e.g. in suspensions of tobacco viruses) \cite{Lettinga2004,Lettinga05}. Of course, these considerations prompt the question how a closed-loop feedback scheme such as the one proposed here could be realized experimentally.

In that context we would like to note that the "control targets" chosen in our study, that is, the components of the order parameter tensor ${\bf a}(t)$, are related to material properties accessible in experiments.
In particular, the components $a_1$ and $a_2$ are related to the so-called flow angle \cite{Hess04} (i.e. the angle between the nematic director
and the flow direction) and to birefringence \cite{Burghardt1998,Kilfoil2000,Cao1993}. Moreover, there is a direct relation between the alignment
and the stress tensor, which can be measured in rheological experiments. This offers an additional route to extract at least some (typically non-diagonal)
components of ${\bf a}$. Given that it can be difficult to monitor all components of ${\bf a}$ at once, the present diagonal control scheme, where all
$a_i$ ($i=0,\dots,4$) are treated on the same footing, may seem too artificial.
We note, however, that the control scheme could be easily modified such that only some, experimentally accessible, components of ${\bf a}$ are controlled
(see Ref.~\cite{Flunkert2011} for an application in a laser with feedback). The only drawback is that an analytical treatment is then impossible.

We further note that {\it additive} terms, such as the control terms in our TDFC scheme, also arise if the system is under the influence of
an external (magnetic or electrical) field. For example, a magnetic field $\bm{H}$ would lead to an additional
free energy contribution of the form  
$\Phi=-\bm{H}\cdot\bm{a}\cdot\bm{H}$, yielding additive terms $\propto H^2$ in the equation of motion [see Eq.~(\ref{eq:HomSyst1})]. One idea could therefore be to choose the strength of the external field, or rather, of $H^2$, to be linearly dependent on the order parameters to generate a linear control scheme such as the one used here. Further possibilities to detect and thus, to control, the orientational motion could arise if the
particles carry electric or magnetic moments on their own, such that time-dependent director motion directly leads to electromagnetic fields. Theoretically, the dynamical properties of such systems (without feedback control) have already been investigated \cite{Grandner07}.

\begin{acknowledgments}
We gratefully acknowledge financial support from the Deutsche
Forschungsgemeinschaft through the Collaborative Research Center~910.
\end{acknowledgments}

\appendix
\section{}
\label{appA}
The components of the vector ${\bm{\Phi}}$ (consists of projections of ${\Phi'}(\bm{a})$ on to the tensor basis)
\begin{align}
  \phi_0= &({\theta}-3a_0+2a^2)a_0+3(a^2_1+a^2_2)-{\frac{3}{2}}(a^2_3+a^2_4)\notag\\
  \phi_1= &({\theta}+6a_0+2a^2)a_1-{\frac{3}{2}}{\sqrt{3}}(a^2_3-a^2_4)\notag\\
  \phi_2= &({\theta}+6a_0+2a^2)a_2-3{\sqrt{3}}a_3a_4\notag\\
  \phi_3= &({\theta}-3a_0+2a^2)a_3-3{\sqrt{3}}(a_1a_3+a_2a_4)\notag\\
  \phi_4= &({\theta}-3a_0+2a^2)a_4-3{\sqrt{3}}(a_2a_3-a_1a_4)\label{eq:HomSyst2}.
\end{align}
See also Refs.~\cite{Hess04,Klapp10,Strehober2013a}.\\

The elements of the Jacobian, $J_{ij}=\partial F_i/\partial q_j$, are given by
\begin{align}
\label{J_original}
J_{00} & = -(\theta-6a_0+4a_0^2+2a^2)\\
J_{01} & = -(4a_0a_1+6a_1)\\
J_{02} & = -(4a_0a_2+6a_2)\\
J_{03} & = -(4a_0a_3-3a_3)\\
J_{04} & = -(4a_0a_4-3a_4),
\end{align}

\begin{align}
J_{10} & =  -((6+4a_0)a_1)\\
J_{11} & =  -(\theta+6a_0+2a^2+4a_1^2)\\
J_{12} & =  -(4a_1a_2)+\dot{{\gamma}}\\
J_{13} & =  -(4a_1a_3-3\sqrt{3}a_3)\\
J_{14} & =  -(4a_1a_4+3\sqrt{3}a_4),
\end{align}

\begin{align}
J_{20} & = -((6+4a_0)a_2)\\
J_{21} & = -(4a_1a_2)-\dot{\gamma}\\
J_{22} & = -(\theta+6a_0+2a^2+4a_2^2)\\
J_{23} & = -(4a_2a_3-3\sqrt{3}a_4)\\
J_{24} & = -(4a_2a_4-3\sqrt{3}a_3),
\end{align}

\begin{align}
J_{30} & =  -((-3+4a_0)a_3)\\
J_{31} & =  -(4a_1a_3-3\sqrt{3}a_3)\\
J_{32} & =  -(4a_2a_3-3\sqrt{3}a_4)\\
J_{33} & =  -(4a_3^2-3\sqrt{3}a_1+(\theta-3a_0+2a^2))\\
J_{34} & =  -(4a_4a_3-3\sqrt{3}a_2)+{\frac{1}{2}}\dot{{\gamma}},
\end{align}

\begin{align}
J_{40} & =  -((-3+4a_0)a_4)\\
J_{41} & =  -(4a_1a_4+3\sqrt{3}a_4)\\
J_{42} & =  -(4a_2a_4-3\sqrt{3}a_3)\\
J_{43} & =  -(4a_3a_4-3\sqrt{3}a_2)-{\frac{1}{2}}\dot{{\gamma}}\\
J_{44} & =  -(4a_4^2+3\sqrt{3}a_1+(\theta-3a_0+2a^2)).
\end{align}

\bibliography{Strehober.bib}

\begin{thebibliography}{64}%
\makeatletter
\providecommand \@ifxundefined [1]{%
 \@ifx{#1\undefined}
}%
\providecommand \@ifnum [1]{%
 \ifnum #1\expandafter \@firstoftwo
 \else \expandafter \@secondoftwo
 \fi
}%
\providecommand \@ifx [1]{%
 \ifx #1\expandafter \@firstoftwo
 \else \expandafter \@secondoftwo
 \fi
}%
\providecommand \natexlab [1]{#1}%
\providecommand \enquote  [1]{``#1''}%
\providecommand \bibnamefont  [1]{#1}%
\providecommand \bibfnamefont [1]{#1}%
\providecommand \citenamefont [1]{#1}%
\providecommand \href@noop [0]{\@secondoftwo}%
\providecommand \href [0]{\begingroup \@sanitize@url \@href}%
\providecommand \@href[1]{\@@startlink{#1}\@@href}%
\providecommand \@@href[1]{\endgroup#1\@@endlink}%
\providecommand \@sanitize@url [0]{\catcode `\\12\catcode `\$12\catcode
  `\&12\catcode `\#12\catcode `\^12\catcode `\_12\catcode `\%12\relax}%
\providecommand \@@startlink[1]{}%
\providecommand \@@endlink[0]{}%
\providecommand \url  [0]{\begingroup\@sanitize@url \@url }%
\providecommand \@url [1]{\endgroup\@href {#1}{\urlprefix }}%
\providecommand \urlprefix  [0]{URL }%
\providecommand \Eprint [0]{\href }%
\providecommand \doibase [0]{http://dx.doi.org/}%
\providecommand \selectlanguage [0]{\@gobble}%
\providecommand \bibinfo  [0]{\@secondoftwo}%
\providecommand \bibfield  [0]{\@secondoftwo}%
\providecommand \translation [1]{[#1]}%
\providecommand \BibitemOpen [0]{}%
\providecommand \bibitemStop [0]{}%
\providecommand \bibitemNoStop [0]{.\EOS\space}%
\providecommand \EOS [0]{\spacefactor3000\relax}%
\providecommand \BibitemShut  [1]{\csname bibitem#1\endcsname}%
\let\auto@bib@innerbib\@empty
\bibitem [{\citenamefont {Rien\"{a}cker}\ \emph
  {et~al.}(2002{\natexlab{a}})\citenamefont {Rien\"{a}cker}, \citenamefont
  {Kr\"{o}ger},\ and\ \citenamefont {Hess}}]{Rien02}%
  \BibitemOpen
  \bibfield  {author} {\bibinfo {author} {\bibfnamefont {G.}~\bibnamefont
  {Rien\"{a}cker}}, \bibinfo {author} {\bibfnamefont {M.}~\bibnamefont
  {Kr\"{o}ger}}, \ and\ \bibinfo {author} {\bibfnamefont {S.}~\bibnamefont
  {Hess}},\ }\href {\doibase 10.1103/PhysRevE.66.040702} {\bibfield  {journal}
  {\bibinfo  {journal} {Phys. Rev. E}\ }\textbf {\bibinfo {volume} {66}},\
  \bibinfo {pages} {040702(R)} (\bibinfo {year}
  {2002}{\natexlab{a}})}\BibitemShut {NoStop}%
\bibitem [{\citenamefont {Rien\"{a}cker}\ \emph
  {et~al.}(2002{\natexlab{b}})\citenamefont {Rien\"{a}cker}, \citenamefont
  {Kr\"{o}ger},\ and\ \citenamefont {Hess}}]{Rien02a}%
  \BibitemOpen
  \bibfield  {author} {\bibinfo {author} {\bibfnamefont {G.}~\bibnamefont
  {Rien\"{a}cker}}, \bibinfo {author} {\bibfnamefont {M.}~\bibnamefont
  {Kr\"{o}ger}}, \ and\ \bibinfo {author} {\bibfnamefont {S.}~\bibnamefont
  {Hess}},\ }\href {\doibase 10.1016/S0378-4371(02)01008-7} {\bibfield
  {journal} {\bibinfo  {journal} {Physica A}\ }\textbf {\bibinfo {volume}
  {315}},\ \bibinfo {pages} {537} (\bibinfo {year}
  {2002}{\natexlab{b}})}\BibitemShut {NoStop}%
\bibitem [{\citenamefont {Grosso}\ \emph {et~al.}(2003)\citenamefont {Grosso},
  \citenamefont {Crescitelli}, \citenamefont {Somma}, \citenamefont {Vermant},
  \citenamefont {Moldenaers},\ and\ \citenamefont {Maffettone}}]{Grosso03}%
  \BibitemOpen
  \bibfield  {author} {\bibinfo {author} {\bibfnamefont {M.}~\bibnamefont
  {Grosso}}, \bibinfo {author} {\bibfnamefont {S.}~\bibnamefont {Crescitelli}},
  \bibinfo {author} {\bibfnamefont {E.}~\bibnamefont {Somma}}, \bibinfo
  {author} {\bibfnamefont {J.}~\bibnamefont {Vermant}}, \bibinfo {author}
  {\bibfnamefont {P.}~\bibnamefont {Moldenaers}}, \ and\ \bibinfo {author}
  {\bibfnamefont {P.~L.}\ \bibnamefont {Maffettone}},\ }\href@noop {}
  {\bibfield  {journal} {\bibinfo  {journal} {Phys.~Rev.~Lett.}\ }\textbf
  {\bibinfo {volume} {90}},\ \bibinfo {pages} {098304} (\bibinfo {year}
  {2003})}\BibitemShut {NoStop}%
\bibitem [{\citenamefont {Forest}\ \emph {et~al.}(2004)\citenamefont {Forest},
  \citenamefont {Wang},\ and\ \citenamefont {Zhou}}]{Forest04}%
  \BibitemOpen
  \bibfield  {author} {\bibinfo {author} {\bibfnamefont {M.~G.}\ \bibnamefont
  {Forest}}, \bibinfo {author} {\bibfnamefont {Q.}~\bibnamefont {Wang}}, \ and\
  \bibinfo {author} {\bibfnamefont {R.}~\bibnamefont {Zhou}},\ }\href@noop {}
  {\bibfield  {journal} {\bibinfo  {journal} {Rheol. Acta}\ }\textbf {\bibinfo
  {volume} {86}},\ \bibinfo {pages} {80} (\bibinfo {year} {2004})}\BibitemShut
  {NoStop}%
\bibitem [{\citenamefont {Ripoll}\ \emph {et~al.}(2008)\citenamefont {Ripoll},
  \citenamefont {Holmqvist}, \citenamefont {Winkler}, \citenamefont {Gompper},
  \citenamefont {Dhont},\ and\ \citenamefont {Lettinga}}]{Ripoll08}%
  \BibitemOpen
  \bibfield  {author} {\bibinfo {author} {\bibfnamefont {M.}~\bibnamefont
  {Ripoll}}, \bibinfo {author} {\bibfnamefont {P.}~\bibnamefont {Holmqvist}},
  \bibinfo {author} {\bibfnamefont {R.~G.}\ \bibnamefont {Winkler}}, \bibinfo
  {author} {\bibfnamefont {G.}~\bibnamefont {Gompper}}, \bibinfo {author}
  {\bibfnamefont {J.~K.~G.}\ \bibnamefont {Dhont}}, \ and\ \bibinfo {author}
  {\bibfnamefont {M.~P.}\ \bibnamefont {Lettinga}},\ }\href@noop {} {\bibfield
  {journal} {\bibinfo  {journal} {Phys. Rev. Lett.}\ }\textbf {\bibinfo
  {volume} {101}},\ \bibinfo {pages} {168302} (\bibinfo {year}
  {2008})}\BibitemShut {NoStop}%
\bibitem [{\citenamefont {Das}\ \emph {et~al.}(2005)\citenamefont {Das},
  \citenamefont {Chakrabarti}, \citenamefont {Dasgupta}, \citenamefont
  {Ramaswamy},\ and\ \citenamefont {Sood}}]{Das05}%
  \BibitemOpen
  \bibfield  {author} {\bibinfo {author} {\bibfnamefont {M.~B.}\ \bibnamefont
  {Das}}, \bibinfo {author} {\bibfnamefont {C.}~\bibnamefont {Chakrabarti}},
  \bibinfo {author} {\bibfnamefont {S.}~\bibnamefont {Dasgupta}}, \bibinfo
  {author} {\bibnamefont {Ramaswamy}}, \ and\ \bibinfo {author} {\bibfnamefont
  {A.~K.}\ \bibnamefont {Sood}},\ }\href@noop {} {\bibfield  {journal}
  {\bibinfo  {journal} {Phys. Rev E}\ }\textbf {\bibinfo {volume} {71}},\
  \bibinfo {pages} {021707} (\bibinfo {year} {2005})}\BibitemShut {NoStop}%
\bibitem [{\citenamefont {Hess}(1975)}]{Hess75}%
  \BibitemOpen
  \bibfield  {author} {\bibinfo {author} {\bibfnamefont {S.}~\bibnamefont
  {Hess}},\ }\href@noop {} {\bibfield  {journal} {\bibinfo  {journal}
  {Z.~Naturforsch.}\ }\textbf {\bibinfo {volume} {30a}},\ \bibinfo {pages}
  {728} (\bibinfo {year} {1975})}\BibitemShut {NoStop}%
\bibitem [{\citenamefont {Hess}(1976)}]{Hess76b}%
  \BibitemOpen
  \bibfield  {author} {\bibinfo {author} {\bibfnamefont {S.}~\bibnamefont
  {Hess}},\ }\href@noop {} {\bibfield  {journal} {\bibinfo  {journal}
  {Z.~Naturforsch.}\ }\textbf {\bibinfo {volume} {31a}},\ \bibinfo {pages}
  {1034} (\bibinfo {year} {1976})}\BibitemShut {NoStop}%
\bibitem [{\citenamefont {Doi}(1980)}]{Doi80}%
  \BibitemOpen
  \bibfield  {author} {\bibinfo {author} {\bibfnamefont {M.}~\bibnamefont
  {Doi}},\ }\href {\doibase 10.1080/00150198008209520} {\bibfield  {journal}
  {\bibinfo  {journal} {Ferroelectrics}\ }\textbf {\bibinfo {volume} {30}},\
  \bibinfo {pages} {247} (\bibinfo {year} {1980})}\BibitemShut {NoStop}%
\bibitem [{\citenamefont {Doi}(1981)}]{Doi81}%
  \BibitemOpen
  \bibfield  {author} {\bibinfo {author} {\bibfnamefont {M.}~\bibnamefont
  {Doi}},\ }\href {\doibase 10.1002/pol.1981.180190205} {\bibfield  {journal}
  {\bibinfo  {journal} {J. Polym. Sci. Polym. Phys. Ed.}\ }\textbf {\bibinfo
  {volume} {19}},\ \bibinfo {pages} {229} (\bibinfo {year} {1981})}\BibitemShut
  {NoStop}%
\bibitem [{\citenamefont {Olmsted}\ and\ \citenamefont
  {Goldbart}(1992)}]{Olmsted92}%
  \BibitemOpen
  \bibfield  {author} {\bibinfo {author} {\bibfnamefont {P.}~\bibnamefont
  {Olmsted}}\ and\ \bibinfo {author} {\bibfnamefont {P.}~\bibnamefont
  {Goldbart}},\ }\href {\doibase 10.1103/PhysRevA.46.4966} {\bibfield
  {journal} {\bibinfo  {journal} {Phys. Rev. A}\ }\textbf {\bibinfo {volume}
  {46}},\ \bibinfo {pages} {4966} (\bibinfo {year} {1992})}\BibitemShut
  {NoStop}%
\bibitem [{\citenamefont {Tao}\ \emph {et~al.}(2005)\citenamefont {Tao},
  \citenamefont {den Otter},\ and\ \citenamefont {Briels}}]{Tao05}%
  \BibitemOpen
  \bibfield  {author} {\bibinfo {author} {\bibfnamefont {Y.-G.}\ \bibnamefont
  {Tao}}, \bibinfo {author} {\bibfnamefont {W.~K.}\ \bibnamefont {den Otter}},
  \ and\ \bibinfo {author} {\bibfnamefont {W.~J.}\ \bibnamefont {Briels}},\
  }\href@noop {} {\bibfield  {journal} {\bibinfo  {journal} {Phys. Rev. Lett.}\
  }\textbf {\bibinfo {volume} {95}},\ \bibinfo {pages} {237802} (\bibinfo
  {year} {2005})}\BibitemShut {NoStop}%
\bibitem [{\citenamefont {Tao}\ \emph {et~al.}(2009)\citenamefont {Tao},
  \citenamefont {den Otter},\ and\ \citenamefont {Briels}}]{Tao09}%
  \BibitemOpen
  \bibfield  {author} {\bibinfo {author} {\bibfnamefont {Y.}~\bibnamefont
  {Tao}}, \bibinfo {author} {\bibfnamefont {W.}~\bibnamefont {den Otter}}, \
  and\ \bibinfo {author} {\bibfnamefont {W.}~\bibnamefont {Briels}},\
  }\href@noop {} {\bibfield  {journal} {\bibinfo  {journal} {Europhys. Lett.}\
  }\textbf {\bibinfo {volume} {86}},\ \bibinfo {pages} {56005} (\bibinfo {year}
  {2009})}\BibitemShut {NoStop}%
\bibitem [{\citenamefont {Germano}\ and\ \citenamefont
  {Schmid}(2005)}]{GERM05}%
  \BibitemOpen
  \bibfield  {author} {\bibinfo {author} {\bibfnamefont {G.}~\bibnamefont
  {Germano}}\ and\ \bibinfo {author} {\bibfnamefont {F.}~\bibnamefont
  {Schmid}},\ }\href@noop {} {\bibfield  {journal} {\bibinfo  {journal}
  {J.~Chem.~Phys.}\ }\textbf {\bibinfo {volume} {123}},\ \bibinfo {pages}
  {214703} (\bibinfo {year} {2005})}\BibitemShut {NoStop}%
\bibitem [{\citenamefont {Bandyopadhyay}\ \emph {et~al.}(2000)\citenamefont
  {Bandyopadhyay}, \citenamefont {Basappa},\ and\ \citenamefont
  {Sood}}]{Sood00}%
  \BibitemOpen
  \bibfield  {author} {\bibinfo {author} {\bibfnamefont {R.}~\bibnamefont
  {Bandyopadhyay}}, \bibinfo {author} {\bibfnamefont {G.}~\bibnamefont
  {Basappa}}, \ and\ \bibinfo {author} {\bibfnamefont {A.~K.}\ \bibnamefont
  {Sood}},\ }\href@noop {} {\bibfield  {journal} {\bibinfo  {journal} {Phys.
  Rev. Lett.}\ }\textbf {\bibinfo {volume} {84}},\ \bibinfo {pages} {2022}
  (\bibinfo {year} {2000})}\BibitemShut {NoStop}%
\bibitem [{\citenamefont {Lettinga}\ \emph {et~al.}(2005)\citenamefont
  {Lettinga}, \citenamefont {Dogic}, \citenamefont {Wang},\ and\ \citenamefont
  {Vermant}}]{Lettinga05}%
  \BibitemOpen
  \bibfield  {author} {\bibinfo {author} {\bibfnamefont {M.~P.}\ \bibnamefont
  {Lettinga}}, \bibinfo {author} {\bibfnamefont {Z.}~\bibnamefont {Dogic}},
  \bibinfo {author} {\bibfnamefont {H.}~\bibnamefont {Wang}}, \ and\ \bibinfo
  {author} {\bibfnamefont {J.}~\bibnamefont {Vermant}},\ }\href {\doibase
  10.1021/la050116e} {\bibfield  {journal} {\bibinfo  {journal} {Langmuir}\
  }\textbf {\bibinfo {volume} {21}},\ \bibinfo {pages} {8048} (\bibinfo {year}
  {2005})}\BibitemShut {NoStop}%
\bibitem [{\citenamefont {Fielding}(2007)}]{FIEL07}%
  \BibitemOpen
  \bibfield  {author} {\bibinfo {author} {\bibfnamefont {S.~M.}\ \bibnamefont
  {Fielding}},\ }\href@noop {} {\bibfield  {journal} {\bibinfo  {journal} {Soft
  Matter}\ }\textbf {\bibinfo {volume} {3}},\ \bibinfo {pages} {1262} (\bibinfo
  {year} {2007})}\BibitemShut {NoStop}%
\bibitem [{\citenamefont {Aradian}\ and\ \citenamefont {Cates}(2005)}]{ARA05}%
  \BibitemOpen
  \bibfield  {author} {\bibinfo {author} {\bibfnamefont {A.}~\bibnamefont
  {Aradian}}\ and\ \bibinfo {author} {\bibfnamefont {M.~E.}\ \bibnamefont
  {Cates}},\ }\href@noop {} {\bibfield  {journal} {\bibinfo  {journal}
  {Europhys. Lett.}\ }\textbf {\bibinfo {volume} {70}},\ \bibinfo {pages} {397}
  (\bibinfo {year} {2005})}\BibitemShut {NoStop}%
\bibitem [{\citenamefont {Cates}\ \emph {et~al.}(2002)\citenamefont {Cates},
  \citenamefont {Head},\ and\ \citenamefont {Ajdari}}]{Cates02}%
  \BibitemOpen
  \bibfield  {author} {\bibinfo {author} {\bibfnamefont {M.~E.}\ \bibnamefont
  {Cates}}, \bibinfo {author} {\bibfnamefont {D.~A.}\ \bibnamefont {Head}}, \
  and\ \bibinfo {author} {\bibfnamefont {A.}~\bibnamefont {Ajdari}},\
  }\href@noop {} {\bibfield  {journal} {\bibinfo  {journal} {Phys. Rev. E}\
  }\textbf {\bibinfo {volume} {66}},\ \bibinfo {pages} {025202(R)} (\bibinfo
  {year} {2002})}\BibitemShut {NoStop}%
\bibitem [{\citenamefont {Aradian}\ and\ \citenamefont
  {Cates}(2006)}]{Aradian06}%
  \BibitemOpen
  \bibfield  {author} {\bibinfo {author} {\bibfnamefont {A.}~\bibnamefont
  {Aradian}}\ and\ \bibinfo {author} {\bibfnamefont {M.~E.}\ \bibnamefont
  {Cates}},\ }\href@noop {} {\bibfield  {journal} {\bibinfo  {journal} {Phys.
  Rev. E}\ }\textbf {\bibinfo {volume} {73}},\ \bibinfo {pages} {041508}
  (\bibinfo {year} {2006})}\BibitemShut {NoStop}%
\bibitem [{\citenamefont {Goddard}\ \emph {et~al.}(2008)\citenamefont
  {Goddard}, \citenamefont {Hess}, \citenamefont {Balanov},\ and\ \citenamefont
  {Hess}}]{GOD08}%
  \BibitemOpen
  \bibfield  {author} {\bibinfo {author} {\bibfnamefont {C.}~\bibnamefont
  {Goddard}}, \bibinfo {author} {\bibfnamefont {O.}~\bibnamefont {Hess}},
  \bibinfo {author} {\bibfnamefont {A.~G.}\ \bibnamefont {Balanov}}, \ and\
  \bibinfo {author} {\bibfnamefont {S.}~\bibnamefont {Hess}},\ }\href@noop {}
  {\bibfield  {journal} {\bibinfo  {journal} {Phys. Rev. E}\ }\textbf {\bibinfo
  {volume} {77}},\ \bibinfo {pages} {026311} (\bibinfo {year}
  {2008})}\BibitemShut {NoStop}%
\bibitem [{\citenamefont {Klapp}\ and\ \citenamefont {Hess}(2010)}]{Klapp10}%
  \BibitemOpen
  \bibfield  {author} {\bibinfo {author} {\bibfnamefont {S.~H.~L.}\
  \bibnamefont {Klapp}}\ and\ \bibinfo {author} {\bibfnamefont
  {S.}~\bibnamefont {Hess}},\ }\href@noop {} {\bibfield  {journal} {\bibinfo
  {journal} {Phys. Rev. E}\ }\textbf {\bibinfo {volume} {81}},\ \bibinfo
  {pages} {051711} (\bibinfo {year} {2010})}\BibitemShut {NoStop}%
\bibitem [{\citenamefont {Heidenreich}\ \emph {et~al.}(2009)\citenamefont
  {Heidenreich}, \citenamefont {Hess},\ and\ \citenamefont
  {Klapp}}]{Heidenreich09}%
  \BibitemOpen
  \bibfield  {author} {\bibinfo {author} {\bibfnamefont {S.}~\bibnamefont
  {Heidenreich}}, \bibinfo {author} {\bibfnamefont {S.}~\bibnamefont {Hess}}, \
  and\ \bibinfo {author} {\bibfnamefont {S.~H.~L.}\ \bibnamefont {Klapp}},\
  }\href@noop {} {\bibfield  {journal} {\bibinfo  {journal} {Phys.~Rev.~Lett.}\
  }\textbf {\bibinfo {volume} {102}},\ \bibinfo {pages} {028301} (\bibinfo
  {year} {2009})}\BibitemShut {NoStop}%
\bibitem [{\citenamefont {Hess}\ and\ \citenamefont
  {Kr{\"o}ger}(2004)}]{Hess04}%
  \BibitemOpen
  \bibfield  {author} {\bibinfo {author} {\bibfnamefont {S.}~\bibnamefont
  {Hess}}\ and\ \bibinfo {author} {\bibfnamefont {M.}~\bibnamefont
  {Kr{\"o}ger}},\ }\href@noop {} {\bibfield  {journal} {\bibinfo  {journal}
  {J.~Phys.: Condensed Matter}\ }\textbf {\bibinfo {volume} {16}},\ \bibinfo
  {pages} {S3835} (\bibinfo {year} {2004})}\BibitemShut {NoStop}%
\bibitem [{\citenamefont {Strehober}\ \emph {et~al.}(2012)\citenamefont
  {Strehober}, \citenamefont {Sch{\"o}ll},\ and\ \citenamefont
  {Klapp}}]{Strehober2012}%
  \BibitemOpen
  \bibfield  {author} {\bibinfo {author} {\bibfnamefont {D.~A.}\ \bibnamefont
  {Strehober}}, \bibinfo {author} {\bibfnamefont {E.}~\bibnamefont
  {Sch{\"o}ll}}, \ and\ \bibinfo {author} {\bibfnamefont {S.~H.~L.}\
  \bibnamefont {Klapp}},\ }\href {\doibase 10.1063/1.4765541} {\bibfield
  {journal} {\bibinfo  {journal} {AIP Conference Proceedings}\ }\textbf
  {\bibinfo {volume} {1493}},\ \bibinfo {pages} {552} (\bibinfo {year}
  {2012})}\BibitemShut {NoStop}%
\bibitem [{\citenamefont {Rien\"{a}cker}(2000)}]{Rienacker2000}%
  \BibitemOpen
  \bibfield  {author} {\bibinfo {author} {\bibfnamefont {G.}~\bibnamefont
  {Rien\"{a}cker}},\ }\emph {\bibinfo {title} {{Orientational Dynamics of
  Nematic Liquid Crystals in a Shear Flow}}},\ \href@noop {} {Ph.D. thesis},\
  \bibinfo  {school} {TU Berlin} (\bibinfo {year} {2000})\BibitemShut {NoStop}%
\bibitem [{\citenamefont {Strehober}\ \emph {et~al.}(2013)\citenamefont
  {Strehober}, \citenamefont {Engel},\ and\ \citenamefont
  {Klapp}}]{Strehober2013a}%
  \BibitemOpen
  \bibfield  {author} {\bibinfo {author} {\bibfnamefont {D.~A.}\ \bibnamefont
  {Strehober}}, \bibinfo {author} {\bibfnamefont {H.}~\bibnamefont {Engel}}, \
  and\ \bibinfo {author} {\bibfnamefont {S.~H.~L.}\ \bibnamefont {Klapp}},\
  }\href {\doibase 10.1103/PhysRevE.88.012505} {\bibfield  {journal} {\bibinfo
  {journal} {Phys. Rev. E}\ }\textbf {\bibinfo {volume} {88}},\ \bibinfo
  {pages} {012505} (\bibinfo {year} {2013})}\BibitemShut {NoStop}%
\bibitem [{\citenamefont {de~Andrade~Lima}\ and\ \citenamefont
  {Rey}(2004)}]{Lima2004}%
  \BibitemOpen
  \bibfield  {author} {\bibinfo {author} {\bibfnamefont {L.~R.~P.}\
  \bibnamefont {de~Andrade~Lima}}\ and\ \bibinfo {author} {\bibfnamefont
  {A.~D.}\ \bibnamefont {Rey}},\ }\href {\doibase 10.1103/PhysRevE.70.011701}
  {\bibfield  {journal} {\bibinfo  {journal} {Phys. Rev. E}\ }\textbf {\bibinfo
  {volume} {70}},\ \bibinfo {pages} {011701} (\bibinfo {year}
  {2004})}\BibitemShut {NoStop}%
\bibitem [{\citenamefont {Tsuji}\ and\ \citenamefont {Rey}(1997)}]{Tsuji1997}%
  \BibitemOpen
  \bibfield  {author} {\bibinfo {author} {\bibfnamefont {T.}~\bibnamefont
  {Tsuji}}\ and\ \bibinfo {author} {\bibfnamefont {A.~D.}\ \bibnamefont
  {Rey}},\ }\href {\doibase http://dx.doi.org/10.1016/S0377-0257(97)00037-2}
  {\bibfield  {journal} {\bibinfo  {journal} {Journal of Non-Newtonian Fluid
  Mechanics}\ }\textbf {\bibinfo {volume} {73}},\ \bibinfo {pages} {127 }
  (\bibinfo {year} {1997})}\BibitemShut {NoStop}%
\bibitem [{\citenamefont {Rey}\ and\ \citenamefont {Denn}(2002)}]{Rey2002}%
  \BibitemOpen
  \bibfield  {author} {\bibinfo {author} {\bibfnamefont {A.~D.}\ \bibnamefont
  {Rey}}\ and\ \bibinfo {author} {\bibfnamefont {M.~M.}\ \bibnamefont {Denn}},\
  }\href {\doibase 10.1146/annurev.fluid.34.082401.191847} {\bibfield
  {journal} {\bibinfo  {journal} {Annual Review of Fluid Mechanics}\ }\textbf
  {\bibinfo {volume} {34}},\ \bibinfo {pages} {233} (\bibinfo {year}
  {2002})}\BibitemShut {NoStop}%
\bibitem [{\citenamefont {Pyragas}(1992)}]{Pyragas92}%
  \BibitemOpen
  \bibfield  {author} {\bibinfo {author} {\bibfnamefont {K.}~\bibnamefont
  {Pyragas}},\ }\href@noop {} {\bibfield  {journal} {\bibinfo  {journal}
  {Phys.~Lett.~A}\ }\textbf {\bibinfo {volume} {170}},\ \bibinfo {pages} {421}
  (\bibinfo {year} {1992})}\BibitemShut {NoStop}%
\bibitem [{\citenamefont {Baba}\ \emph {et~al.}(2002)\citenamefont {Baba},
  \citenamefont {Amann}, \citenamefont {Sch{\"o}ll},\ and\ \citenamefont
  {Just}}]{BAB02}%
  \BibitemOpen
  \bibfield  {author} {\bibinfo {author} {\bibfnamefont {N.}~\bibnamefont
  {Baba}}, \bibinfo {author} {\bibfnamefont {A.}~\bibnamefont {Amann}},
  \bibinfo {author} {\bibfnamefont {E.}~\bibnamefont {Sch{\"o}ll}}, \ and\
  \bibinfo {author} {\bibfnamefont {W.}~\bibnamefont {Just}},\ }\href@noop {}
  {\bibfield  {journal} {\bibinfo  {journal} {Phys.~Rev.~Lett.}\ }\textbf
  {\bibinfo {volume} {89}},\ \bibinfo {pages} {074101} (\bibinfo {year}
  {2002})}\BibitemShut {NoStop}%
\bibitem [{\citenamefont {Unkelbach}\ \emph {et~al.}(2003)\citenamefont
  {Unkelbach}, \citenamefont {Amann}, \citenamefont {Just},\ and\ \citenamefont
  {Sch{\"o}ll}}]{UNK03}%
  \BibitemOpen
  \bibfield  {author} {\bibinfo {author} {\bibfnamefont {J.}~\bibnamefont
  {Unkelbach}}, \bibinfo {author} {\bibfnamefont {A.}~\bibnamefont {Amann}},
  \bibinfo {author} {\bibfnamefont {W.}~\bibnamefont {Just}}, \ and\ \bibinfo
  {author} {\bibfnamefont {E.}~\bibnamefont {Sch{\"o}ll}},\ }\href@noop {}
  {\bibfield  {journal} {\bibinfo  {journal} {Phys.~Rev.~E}\ }\textbf {\bibinfo
  {volume} {68}},\ \bibinfo {pages} {026204} (\bibinfo {year}
  {2003})}\BibitemShut {NoStop}%
\bibitem [{\citenamefont {Schlesner}\ \emph {et~al.}(2003)\citenamefont
  {Schlesner}, \citenamefont {Amann}, \citenamefont {Janson}, \citenamefont
  {Just},\ and\ \citenamefont {Sch{\"o}ll}}]{SCH03a}%
  \BibitemOpen
  \bibfield  {author} {\bibinfo {author} {\bibfnamefont {J.}~\bibnamefont
  {Schlesner}}, \bibinfo {author} {\bibfnamefont {A.}~\bibnamefont {Amann}},
  \bibinfo {author} {\bibfnamefont {N.~B.}\ \bibnamefont {Janson}}, \bibinfo
  {author} {\bibfnamefont {W.}~\bibnamefont {Just}}, \ and\ \bibinfo {author}
  {\bibfnamefont {E.}~\bibnamefont {Sch{\"o}ll}},\ }\href@noop {} {\bibfield
  {journal} {\bibinfo  {journal} {Phys.~Rev.~E}\ }\textbf {\bibinfo {volume}
  {68}},\ \bibinfo {pages} {066208} (\bibinfo {year} {2003})}\BibitemShut
  {NoStop}%
\bibitem [{\citenamefont {Kehrt}\ \emph {et~al.}(2009)\citenamefont {Kehrt},
  \citenamefont {H{\"o}vel}, \citenamefont {Flunkert}, \citenamefont {Dahlem},
  \citenamefont {Rodin},\ and\ \citenamefont {Sch{\"o}ll}}]{KEH09}%
  \BibitemOpen
  \bibfield  {author} {\bibinfo {author} {\bibfnamefont {M.}~\bibnamefont
  {Kehrt}}, \bibinfo {author} {\bibfnamefont {P.}~\bibnamefont {H{\"o}vel}},
  \bibinfo {author} {\bibfnamefont {V.}~\bibnamefont {Flunkert}}, \bibinfo
  {author} {\bibfnamefont {M.~A.}\ \bibnamefont {Dahlem}}, \bibinfo {author}
  {\bibfnamefont {P.}~\bibnamefont {Rodin}}, \ and\ \bibinfo {author}
  {\bibfnamefont {E.}~\bibnamefont {Sch{\"o}ll}},\ }\href@noop {} {\bibfield
  {journal} {\bibinfo  {journal} {Eur. Phys. J. B}\ }\textbf {\bibinfo {volume}
  {68}},\ \bibinfo {pages} {557} (\bibinfo {year} {2009})}\BibitemShut
  {NoStop}%
\bibitem [{\citenamefont {Dahms}\ \emph {et~al.}(2008)\citenamefont {Dahms},
  \citenamefont {H{\"o}vel},\ and\ \citenamefont {Sch{\"o}ll}}]{DAH08b}%
  \BibitemOpen
  \bibfield  {author} {\bibinfo {author} {\bibfnamefont {T.}~\bibnamefont
  {Dahms}}, \bibinfo {author} {\bibfnamefont {P.}~\bibnamefont {H{\"o}vel}}, \
  and\ \bibinfo {author} {\bibfnamefont {E.}~\bibnamefont {Sch{\"o}ll}},\
  }\href@noop {} {\bibfield  {journal} {\bibinfo  {journal} {Phys.~Rev.~E}\
  }\textbf {\bibinfo {volume} {78}},\ \bibinfo {pages} {056213} (\bibinfo
  {year} {2008})}\BibitemShut {NoStop}%
\bibitem [{\citenamefont {Dahms}\ \emph {et~al.}(2010)\citenamefont {Dahms},
  \citenamefont {Flunkert}, \citenamefont {Henneberger}, \citenamefont
  {H{\"o}vel}, \citenamefont {Schikora}, \citenamefont {Sch{\"o}ll},\ and\
  \citenamefont {W{\"u}nsche}}]{DAH10}%
  \BibitemOpen
  \bibfield  {author} {\bibinfo {author} {\bibfnamefont {T.}~\bibnamefont
  {Dahms}}, \bibinfo {author} {\bibfnamefont {V.}~\bibnamefont {Flunkert}},
  \bibinfo {author} {\bibfnamefont {F.}~\bibnamefont {Henneberger}}, \bibinfo
  {author} {\bibfnamefont {P.}~\bibnamefont {H{\"o}vel}}, \bibinfo {author}
  {\bibfnamefont {S.}~\bibnamefont {Schikora}}, \bibinfo {author}
  {\bibfnamefont {E.}~\bibnamefont {Sch{\"o}ll}}, \ and\ \bibinfo {author}
  {\bibfnamefont {H.~J.}\ \bibnamefont {W{\"u}nsche}},\ }\href@noop {}
  {\bibfield  {journal} {\bibinfo  {journal} {Eur. Phys.~J.~ST}\ }\textbf
  {\bibinfo {volume} {191}},\ \bibinfo {pages} {71} (\bibinfo {year}
  {2010})}\BibitemShut {NoStop}%
\bibitem [{\citenamefont {Schlesner}\ \emph {et~al.}(2006)\citenamefont
  {Schlesner}, \citenamefont {Zykov}, \citenamefont {Engel},\ and\
  \citenamefont {Sch{\"o}ll}}]{SCH06c}%
  \BibitemOpen
  \bibfield  {author} {\bibinfo {author} {\bibfnamefont {J.}~\bibnamefont
  {Schlesner}}, \bibinfo {author} {\bibfnamefont {V.}~\bibnamefont {Zykov}},
  \bibinfo {author} {\bibfnamefont {H.}~\bibnamefont {Engel}}, \ and\ \bibinfo
  {author} {\bibfnamefont {E.}~\bibnamefont {Sch{\"o}ll}},\ }\href@noop {}
  {\bibfield  {journal} {\bibinfo  {journal} {Phys.~Rev.~E}\ }\textbf {\bibinfo
  {volume} {74}},\ \bibinfo {pages} {046215} (\bibinfo {year}
  {2006})}\BibitemShut {NoStop}%
\bibitem [{\citenamefont {Schlesner}\ \emph {et~al.}(2008)\citenamefont
  {Schlesner}, \citenamefont {Zykov}, \citenamefont {Brandst{\"a}dter},
  \citenamefont {Gerdes},\ and\ \citenamefont {Engel}}]{Schlesner08}%
  \BibitemOpen
  \bibfield  {author} {\bibinfo {author} {\bibfnamefont {J.}~\bibnamefont
  {Schlesner}}, \bibinfo {author} {\bibfnamefont {V.~S.}\ \bibnamefont
  {Zykov}}, \bibinfo {author} {\bibfnamefont {H.}~\bibnamefont
  {Brandst{\"a}dter}}, \bibinfo {author} {\bibfnamefont {I.}~\bibnamefont
  {Gerdes}}, \ and\ \bibinfo {author} {\bibfnamefont {H.}~\bibnamefont
  {Engel}},\ }\href@noop {} {\bibfield  {journal} {\bibinfo  {journal} {New
  J.~Phys.}\ }\textbf {\bibinfo {volume} {10}},\ \bibinfo {pages} {015003}
  (\bibinfo {year} {2008})}\BibitemShut {NoStop}%
\bibitem [{\citenamefont {Kyrychko}\ \emph {et~al.}(2009)\citenamefont
  {Kyrychko}, \citenamefont {Blyuss}, \citenamefont {Hogan},\ and\
  \citenamefont {Sch{\"o}ll}}]{KYR09}%
  \BibitemOpen
  \bibfield  {author} {\bibinfo {author} {\bibfnamefont {Y.~N.}\ \bibnamefont
  {Kyrychko}}, \bibinfo {author} {\bibfnamefont {K.~B.}\ \bibnamefont
  {Blyuss}}, \bibinfo {author} {\bibfnamefont {S.~J.}\ \bibnamefont {Hogan}}, \
  and\ \bibinfo {author} {\bibfnamefont {E.}~\bibnamefont {Sch{\"o}ll}},\
  }\href@noop {} {\bibfield  {journal} {\bibinfo  {journal} {Chaos}\ }\textbf
  {\bibinfo {volume} {19}},\ \bibinfo {pages} {043126} (\bibinfo {year}
  {2009})}\BibitemShut {NoStop}%
\bibitem [{\citenamefont {Dahlem}\ \emph {et~al.}(2008)\citenamefont {Dahlem},
  \citenamefont {Schneider},\ and\ \citenamefont {Sch{\"o}ll}}]{DAH08}%
  \BibitemOpen
  \bibfield  {author} {\bibinfo {author} {\bibfnamefont {M.~A.}\ \bibnamefont
  {Dahlem}}, \bibinfo {author} {\bibfnamefont {F.~M.}\ \bibnamefont
  {Schneider}}, \ and\ \bibinfo {author} {\bibfnamefont {E.}~\bibnamefont
  {Sch{\"o}ll}},\ }\href@noop {} {\bibfield  {journal} {\bibinfo  {journal}
  {Chaos}\ }\textbf {\bibinfo {volume} {18}},\ \bibinfo {pages} {026110}
  (\bibinfo {year} {2008})}\BibitemShut {NoStop}%
\bibitem [{\citenamefont {Sch{\"o}ll}\ \emph {et~al.}(2009)\citenamefont
  {Sch{\"o}ll}, \citenamefont {Hiller}, \citenamefont {H{\"o}vel},\ and\
  \citenamefont {Dahlem}}]{SCH08}%
  \BibitemOpen
  \bibfield  {author} {\bibinfo {author} {\bibfnamefont {E.}~\bibnamefont
  {Sch{\"o}ll}}, \bibinfo {author} {\bibfnamefont {G.}~\bibnamefont {Hiller}},
  \bibinfo {author} {\bibfnamefont {P.}~\bibnamefont {H{\"o}vel}}, \ and\
  \bibinfo {author} {\bibfnamefont {M.~A.}\ \bibnamefont {Dahlem}},\
  }\href@noop {} {\bibfield  {journal} {\bibinfo  {journal} {Phil. Trans.~R.
  Soc.~A}\ }\textbf {\bibinfo {volume} {367}},\ \bibinfo {pages} {1079}
  (\bibinfo {year} {2009})}\BibitemShut {NoStop}%
\bibitem [{\citenamefont {Panchuk}\ \emph {et~al.}(2012)\citenamefont
  {Panchuk}, \citenamefont {Rosin}, \citenamefont {H{\"o}vel},\ and\
  \citenamefont {Sch{\"o}ll}}]{PAN12}%
  \BibitemOpen
  \bibfield  {author} {\bibinfo {author} {\bibfnamefont {A.}~\bibnamefont
  {Panchuk}}, \bibinfo {author} {\bibfnamefont {D.~P.}\ \bibnamefont {Rosin}},
  \bibinfo {author} {\bibfnamefont {P.}~\bibnamefont {H{\"o}vel}}, \ and\
  \bibinfo {author} {\bibfnamefont {E.}~\bibnamefont {Sch{\"o}ll}},\
  }\href@noop {} {\bibfield  {journal} {\bibinfo  {journal} {Int. J.~Bif.
  Chaos}\ } (\bibinfo {year} {2012})}\BibitemShut {NoStop}%
\bibitem [{\citenamefont {Sch{\"o}ll}\ and\ \citenamefont
  {Schuster}(2008)}]{Schoellbuch}%
  \BibitemOpen
  \bibinfo {editor} {\bibfnamefont {E.}~\bibnamefont {Sch{\"o}ll}}\ and\
  \bibinfo {editor} {\bibfnamefont {H.~G.}\ \bibnamefont {Schuster}},\ eds.,\
  \href@noop {} {\emph {\bibinfo {title} {{Handbook of Chaos Control}}}}\
  (\bibinfo  {publisher} {Wiley VCH},\ \bibinfo {address} {Weinheim},\ \bibinfo
  {year} {2008})\BibitemShut {NoStop}%
\bibitem [{\citenamefont {Sch{\"o}ll}(2009)}]{Schoellreview}%
  \BibitemOpen
  \bibfield  {author} {\bibinfo {author} {\bibfnamefont {E.}~\bibnamefont
  {Sch{\"o}ll}},\ }\href@noop {} {\emph {\bibinfo {title} {Nonlinear Dynamics
  of Nanosystems}}},\ edited by\ \bibinfo {editor} {\bibfnamefont
  {G.}~\bibnamefont {Radons}}, \bibinfo {editor} {\bibfnamefont
  {B.}~\bibnamefont {Rumpf}}, \ and\ \bibinfo {editor} {\bibfnamefont {H.~G.}\
  \bibnamefont {Schuster}}\ (\bibinfo  {publisher} {Wiley-VCH, Weinheim},\
  \bibinfo {year} {2009})\BibitemShut {NoStop}%
\bibitem [{\citenamefont {H{\"o}vel}\ and\ \citenamefont
  {Sch{\"o}ll}(2005)}]{HOV05}%
  \BibitemOpen
  \bibfield  {author} {\bibinfo {author} {\bibfnamefont {P.}~\bibnamefont
  {H{\"o}vel}}\ and\ \bibinfo {author} {\bibfnamefont {E.}~\bibnamefont
  {Sch{\"o}ll}},\ }\href@noop {} {\bibfield  {journal} {\bibinfo  {journal}
  {Phys.~Rev.~E}\ }\textbf {\bibinfo {volume} {72}},\ \bibinfo {pages} {046203}
  (\bibinfo {year} {2005})}\BibitemShut {NoStop}%
\bibitem [{\citenamefont {Sonnet}\ \emph {et~al.}(1995)\citenamefont {Sonnet},
  \citenamefont {Kilian},\ and\ \citenamefont {Hess}}]{Sonnet1995}%
  \BibitemOpen
  \bibfield  {author} {\bibinfo {author} {\bibfnamefont {A.}~\bibnamefont
  {Sonnet}}, \bibinfo {author} {\bibfnamefont {A.}~\bibnamefont {Kilian}}, \
  and\ \bibinfo {author} {\bibfnamefont {S.}~\bibnamefont {Hess}},\ }\href
  {\doibase 10.1103/PhysRevE.52.718} {\bibfield  {journal} {\bibinfo  {journal}
  {Phys. Rev. E}\ }\textbf {\bibinfo {volume} {52}},\ \bibinfo {pages} {718}
  (\bibinfo {year} {1995})}\BibitemShut {NoStop}%
\bibitem [{\citenamefont {Grandner}\ \emph
  {et~al.}(2007{\natexlab{a}})\citenamefont {Grandner}, \citenamefont
  {Heidenreich}, \citenamefont {Hess},\ and\ \citenamefont
  {Klapp}}]{Grandner07}%
  \BibitemOpen
  \bibfield  {author} {\bibinfo {author} {\bibfnamefont {S.}~\bibnamefont
  {Grandner}}, \bibinfo {author} {\bibfnamefont {S.}~\bibnamefont
  {Heidenreich}}, \bibinfo {author} {\bibfnamefont {S.}~\bibnamefont {Hess}}, \
  and\ \bibinfo {author} {\bibfnamefont {S.~H.~L.}\ \bibnamefont {Klapp}},\
  }\href@noop {} {\bibfield  {journal} {\bibinfo  {journal} {Eur.~Phys.~J.~E}\
  }\textbf {\bibinfo {volume} {24}},\ \bibinfo {pages} {353} (\bibinfo {year}
  {2007}{\natexlab{a}})}\BibitemShut {NoStop}%
\bibitem [{\citenamefont {Hess}\ and\ \citenamefont
  {Kr\"{o}ger}(2005)}]{Hess2004}%
  \BibitemOpen
  \bibfield  {author} {\bibinfo {author} {\bibfnamefont {S.}~\bibnamefont
  {Hess}}\ and\ \bibinfo {author} {\bibfnamefont {M.}~\bibnamefont
  {Kr\"{o}ger}},\ }in\ \href {\doibase 10.1007/1-4020-2760-5_14} {\emph
  {\bibinfo {booktitle} {Computer Simulations of Liquid Crystals and
  Polymers}}},\ \bibinfo {series} {NATO Science Series II: Mathematics, Physics
  and Chemistry}, Vol.\ \bibinfo {volume} {177},\ \bibinfo {editor} {edited by\
  \bibinfo {editor} {\bibfnamefont {P.}~\bibnamefont {Pasini}}, \bibinfo
  {editor} {\bibfnamefont {C.}~\bibnamefont {Zannoni}}, \ and\ \bibinfo
  {editor} {\bibfnamefont {S.}~\bibnamefont {Zumer}}}\ (\bibinfo  {publisher}
  {Springer Netherlands},\ \bibinfo {year} {2005})\ pp.\ \bibinfo {pages}
  {295--333}\BibitemShut {NoStop}%
\bibitem [{\citenamefont {Grandner}\ \emph
  {et~al.}(2007{\natexlab{b}})\citenamefont {Grandner}, \citenamefont
  {Heidenreich}, \citenamefont {Hess},\ and\ \citenamefont
  {Klapp}}]{GrandnerEPJ2007}%
  \BibitemOpen
  \bibfield  {author} {\bibinfo {author} {\bibfnamefont {S.}~\bibnamefont
  {Grandner}}, \bibinfo {author} {\bibfnamefont {S.}~\bibnamefont
  {Heidenreich}}, \bibinfo {author} {\bibfnamefont {S.}~\bibnamefont {Hess}}, \
  and\ \bibinfo {author} {\bibfnamefont {S.~H.~L.}\ \bibnamefont {Klapp}},\
  }\href {\doibase 10.1140/epje/i2007-10246-8} {\bibfield  {journal} {\bibinfo
  {journal} {Eur. Phys. J. E}\ }\textbf {\bibinfo {volume} {24}},\ \bibinfo
  {pages} {353} (\bibinfo {year} {2007}{\natexlab{b}})}\BibitemShut {NoStop}%
\bibitem [{\citenamefont {Grandner}\ \emph
  {et~al.}(2007{\natexlab{c}})\citenamefont {Grandner}, \citenamefont
  {Heidenreich}, \citenamefont {Ilg}, \citenamefont {Klapp},\ and\
  \citenamefont {Hess}}]{GrandnerPRE2007}%
  \BibitemOpen
  \bibfield  {author} {\bibinfo {author} {\bibfnamefont {S.}~\bibnamefont
  {Grandner}}, \bibinfo {author} {\bibfnamefont {S.}~\bibnamefont
  {Heidenreich}}, \bibinfo {author} {\bibfnamefont {P.}~\bibnamefont {Ilg}},
  \bibinfo {author} {\bibfnamefont {S.}~\bibnamefont {Klapp}}, \ and\ \bibinfo
  {author} {\bibfnamefont {S.}~\bibnamefont {Hess}},\ }\href {\doibase
  10.1103/PhysRevE.75.040701} {\bibfield  {journal} {\bibinfo  {journal} {Phys.
  Rev. E}\ }\textbf {\bibinfo {volume} {75}},\ \bibinfo {pages} {2} (\bibinfo
  {year} {2007}{\natexlab{c}})}\BibitemShut {NoStop}%
\bibitem [{\citenamefont {de~Gennes}\ and\ \citenamefont
  {Prost}(1993)}]{deGennes}%
  \BibitemOpen
  \bibfield  {author} {\bibinfo {author} {\bibfnamefont {P.~G.}\ \bibnamefont
  {de~Gennes}}\ and\ \bibinfo {author} {\bibfnamefont {J.}~\bibnamefont
  {Prost}},\ }\href@noop {} {\emph {\bibinfo {title} {{The physics of liquid
  crystals}}}}\ (\bibinfo  {publisher} {Clarendon},\ \bibinfo {address}
  {Oxford},\ \bibinfo {year} {1993})\BibitemShut {NoStop}%
\bibitem [{\citenamefont {Heidenreich}(2008)}]{Heidenreich2008Thesis}%
  \BibitemOpen
  \bibfield  {author} {\bibinfo {author} {\bibfnamefont {S.}~\bibnamefont
  {Heidenreich}},\ }\emph {\bibinfo {title} {{Orientational Dynamics and Flow
  Properties of Polar and Non-Polar Hard-Rod Fluids}}},\ \href
  {http://opus.kobv.de/tuberlin/volltexte/2009/2146/} {Ph.D. thesis},\ \bibinfo
   {school} {TU Berlin} (\bibinfo {year} {2008})\BibitemShut {NoStop}%
\bibitem [{\citenamefont {Larson}\ and\ \citenamefont
  {{\"O}ttinger}(1991)}]{Larson1991}%
  \BibitemOpen
  \bibfield  {author} {\bibinfo {author} {\bibfnamefont {R.~G.}\ \bibnamefont
  {Larson}}\ and\ \bibinfo {author} {\bibfnamefont {H.~C.}\ \bibnamefont
  {{\"O}ttinger}},\ }\href@noop {} {\bibfield  {journal} {\bibinfo  {journal}
  {Macromolecules}\ }\textbf {\bibinfo {volume} {24}},\ \bibinfo {pages} {6270}
  (\bibinfo {year} {1991})}\BibitemShut {NoStop}%
\bibitem [{\citenamefont {Corless}\ \emph {et~al.}(1996)\citenamefont
  {Corless}, \citenamefont {Gonnet}, \citenamefont {Hare}, \citenamefont
  {Jeffrey},\ and\ \citenamefont {Knuth}}]{Corless96}%
  \BibitemOpen
  \bibfield  {author} {\bibinfo {author} {\bibfnamefont {R.~M.}\ \bibnamefont
  {Corless}}, \bibinfo {author} {\bibfnamefont {G.~H.}\ \bibnamefont {Gonnet}},
  \bibinfo {author} {\bibfnamefont {D.~E.~G.}\ \bibnamefont {Hare}}, \bibinfo
  {author} {\bibfnamefont {D.~J.}\ \bibnamefont {Jeffrey}}, \ and\ \bibinfo
  {author} {\bibfnamefont {D.~E.}\ \bibnamefont {Knuth}},\ }\href@noop {}
  {\bibfield  {journal} {\bibinfo  {journal} {Adv. Comput. Math.}\ }\textbf
  {\bibinfo {volume} {5}},\ \bibinfo {pages} {329} (\bibinfo {year}
  {1996})}\BibitemShut {NoStop}%
\bibitem [{\citenamefont {Yanchuk}\ \emph {et~al.}(2006)\citenamefont
  {Yanchuk}, \citenamefont {Wolfrum}, \citenamefont {H{\"o}vel},\ and\
  \citenamefont {Sch{\"o}ll}}]{YAN06}%
  \BibitemOpen
  \bibfield  {author} {\bibinfo {author} {\bibfnamefont {S.}~\bibnamefont
  {Yanchuk}}, \bibinfo {author} {\bibfnamefont {M.}~\bibnamefont {Wolfrum}},
  \bibinfo {author} {\bibfnamefont {P.}~\bibnamefont {H{\"o}vel}}, \ and\
  \bibinfo {author} {\bibfnamefont {E.}~\bibnamefont {Sch{\"o}ll}},\ }\href
  {\doibase 10.1103/PhysRevE.74.026201} {\bibfield  {journal} {\bibinfo
  {journal} {Phys. Rev. E}\ }\textbf {\bibinfo {volume} {74}},\ \bibinfo
  {pages} {026201} (\bibinfo {year} {2006})}\BibitemShut {NoStop}%
\bibitem [{\citenamefont {Wolfrum}\ \emph {et~al.}(2010)\citenamefont
  {Wolfrum}, \citenamefont {Yanchuk}, \citenamefont {H{\"o}vel},\ and\
  \citenamefont {Sch{\"o}ll}}]{WOL10}%
  \BibitemOpen
  \bibfield  {author} {\bibinfo {author} {\bibfnamefont {M.}~\bibnamefont
  {Wolfrum}}, \bibinfo {author} {\bibfnamefont {S.}~\bibnamefont {Yanchuk}},
  \bibinfo {author} {\bibfnamefont {P.}~\bibnamefont {H{\"o}vel}}, \ and\
  \bibinfo {author} {\bibfnamefont {E.}~\bibnamefont {Sch{\"o}ll}},\ }\href
  {\doibase 10.1140/epjst/e2010-01343-7} {\bibfield  {journal} {\bibinfo
  {journal} {Eur. Phys. J. Special Topics}\ }\textbf {\bibinfo {volume}
  {191}},\ \bibinfo {pages} {91} (\bibinfo {year} {2010})}\BibitemShut
  {NoStop}%
\bibitem [{\citenamefont {Lehnert}\ \emph {et~al.}(2011)\citenamefont
  {Lehnert}, \citenamefont {H{\"o}vel}, \citenamefont {Flunkert}, \citenamefont
  {Guzenko}, \citenamefont {Fradkov},\ and\ \citenamefont
  {Sch{\"o}ll}}]{Lehnert2011}%
  \BibitemOpen
  \bibfield  {author} {\bibinfo {author} {\bibfnamefont {J.}~\bibnamefont
  {Lehnert}}, \bibinfo {author} {\bibfnamefont {P.}~\bibnamefont {H{\"o}vel}},
  \bibinfo {author} {\bibfnamefont {V.}~\bibnamefont {Flunkert}}, \bibinfo
  {author} {\bibfnamefont {P.~Y.}\ \bibnamefont {Guzenko}}, \bibinfo {author}
  {\bibfnamefont {A.~L.}\ \bibnamefont {Fradkov}}, \ and\ \bibinfo {author}
  {\bibfnamefont {E.}~\bibnamefont {Sch{\"o}ll}},\ }\href {\doibase
  10.1063/1.3647320} {\bibfield  {journal} {\bibinfo  {journal} {Chaos}\
  }\textbf {\bibinfo {volume} {21}},\ \bibinfo {eid} {043111} (\bibinfo {year}
  {2011})}\BibitemShut {NoStop}%
\bibitem [{\citenamefont {H{\"o}vel}\ \emph {et~al.}(2010)\citenamefont
  {H{\"o}vel}, \citenamefont {Dahlem},\ and\ \citenamefont
  {Sch{\"o}ll}}]{HOEV10}%
  \BibitemOpen
  \bibfield  {author} {\bibinfo {author} {\bibfnamefont {P.}~\bibnamefont
  {H{\"o}vel}}, \bibinfo {author} {\bibfnamefont {M.~A.}\ \bibnamefont
  {Dahlem}}, \ and\ \bibinfo {author} {\bibfnamefont {E.}~\bibnamefont
  {Sch{\"o}ll}},\ }\href@noop {} {\bibfield  {journal} {\bibinfo  {journal}
  {Int. J. Bifur. Chaos}\ }\textbf {\bibinfo {volume} {20}},\ \bibinfo {pages}
  {813} (\bibinfo {year} {2010})}\BibitemShut {NoStop}%
\bibitem [{\citenamefont {Lettinga}\ and\ \citenamefont
  {Dhont}(2004)}]{Lettinga2004}%
  \BibitemOpen
  \bibfield  {author} {\bibinfo {author} {\bibfnamefont {M.~P.}\ \bibnamefont
  {Lettinga}}\ and\ \bibinfo {author} {\bibfnamefont {J.~K.~G.}\ \bibnamefont
  {Dhont}},\ }\href {\doibase 10.1088/0953-8984/16/38/011} {\bibfield
  {journal} {\bibinfo  {journal} {J. Phys.: Condens. Matter}\ }\textbf
  {\bibinfo {volume} {16}},\ \bibinfo {pages} {S3929} (\bibinfo {year}
  {2004})}\BibitemShut {NoStop}%
\bibitem [{\citenamefont {Burghardt}(1998)}]{Burghardt1998}%
  \BibitemOpen
  \bibfield  {author} {\bibinfo {author} {\bibfnamefont {W.~R.}\ \bibnamefont
  {Burghardt}},\ }\href@noop {} {\bibfield  {journal} {\bibinfo  {journal}
  {Macromolecular Chemistry and Physics}\ }\textbf {\bibinfo {volume} {199}},\
  \bibinfo {pages} {471} (\bibinfo {year} {1998})}\BibitemShut {NoStop}%
\bibitem [{\citenamefont {Kilfoil}\ and\ \citenamefont
  {Callaghan}(2000)}]{Kilfoil2000}%
  \BibitemOpen
  \bibfield  {author} {\bibinfo {author} {\bibfnamefont {M.~L.}\ \bibnamefont
  {Kilfoil}}\ and\ \bibinfo {author} {\bibfnamefont {P.~T.}\ \bibnamefont
  {Callaghan}},\ }\href {\doibase 10.1021/ma000554m} {\bibfield  {journal}
  {\bibinfo  {journal} {Macromolecules}\ }\textbf {\bibinfo {volume} {33}},\
  \bibinfo {pages} {6828} (\bibinfo {year} {2000})}\BibitemShut {NoStop}%
\bibitem [{\citenamefont {Cao}\ and\ \citenamefont {Berne}(1993)}]{Cao1993}%
  \BibitemOpen
  \bibfield  {author} {\bibinfo {author} {\bibfnamefont {J.}~\bibnamefont
  {Cao}}\ and\ \bibinfo {author} {\bibfnamefont {B.~J.}\ \bibnamefont
  {Berne}},\ }\href@noop {} {\bibfield  {journal} {\bibinfo  {journal} {J.
  Chem. Phys.}\ }\textbf {\bibinfo {volume} {99}},\ \bibinfo {pages} {2213}
  (\bibinfo {year} {1993})}\BibitemShut {NoStop}%
\bibitem [{\citenamefont {Flunkert}\ and\ \citenamefont
  {Sch{\"o}ll}(2011)}]{Flunkert2011}%
  \BibitemOpen
  \bibfield  {author} {\bibinfo {author} {\bibfnamefont {V.}~\bibnamefont
  {Flunkert}}\ and\ \bibinfo {author} {\bibfnamefont {E.}~\bibnamefont
  {Sch{\"o}ll}},\ }\href {\doibase 10.1103/PhysRevE.84.016214} {\bibfield
  {journal} {\bibinfo  {journal} {Phys. Rev. E}\ }\textbf {\bibinfo {volume}
  {84}},\ \bibinfo {pages} {016214} (\bibinfo {year} {2011})}\BibitemShut
  {NoStop}%
\end{thebibliography}%

\end{document}